\documentclass[twocolumn]{aastex631}

\usepackage{xcolor}
\usepackage{xspace}
\usepackage{amsmath}
\usepackage{natbib}
\usepackage{bm}

\newcommand{\sersic}[0]{S{\'e}rsic\xspace}
\newcommand{\sphot}[0]{\texttt{SPHOT}\xspace}
\newcommand{\bfit}[1]{\textbf{\textit{#1}}} 
\shorttitle{$R_V$ in an SN~I$\mathrm{a}$ Host from Background Galaxies}
\shortauthors{Murakami et al.}

\graphicspath{{./}{figures/}}

\begin{document}

\title{The Extinction Law in SN Ia Hosts from Background Galaxy Measurements:\\
Toward a 1\% Determination of $H_0$}

\correspondingauthor{Yukei S. Murakami}
\email{ymuraka2@jhu.edu}

\author[0000-0002-8342-3804]{Yukei S. Murakami}
\affiliation{Department of Physics and Astronomy, Johns Hopkins University, Baltimore, MD 21218, USA}

\author[0000-0002-6124-1196]{Adam G. Riess}
\affiliation{Department of Physics and Astronomy, Johns Hopkins University, Baltimore, MD 21218, USA}
\affiliation{Space Telescope Science Institute, 3700 San Martin Drive, Baltimore, MD 21218, USA}

\author[0000-0001-7113-2738]{Henry C. Ferguson}
\affiliation{Space Telescope Science Institute, 3700 San Martin Drive, Baltimore, MD 21218, USA}

\author[0000-0003-3460-0103]{Alexei V. Filippenko}
\affiliation{Department of Astronomy, University of California, Berkeley, CA 94720-3411, USA}

\author[0000-0001-5955-2502]{Thomas~G.~Brink}
\affiliation{Department of Astronomy, University of California, Berkeley, CA 94720-3411, USA}

\author[0000-0002-2636-6508]{WeiKang Zheng}
\affiliation{Department of Astronomy, University of California, Berkeley, CA 94720-3411, USA}

\author[0000-0002-4934-5849]{Dan M. Scolnic}
\affiliation{Department of Physics, Duke University, Durham, NC 27708, USA}

\begin{abstract}
In the most precise distance ladder determination of  $H_0$, the observed near-infrared (NIR) fluxes of Cepheids are corrected for dust, assuming that the extinction law in large, star-forming spiral hosts of Type Ia supernovae (SN Ia) is similar to the Milky Way's average value  of $R_V \approx 3.1$. Intriguingly, studies of SNe~Ia often point to lower values for their hosts ($R_V \lesssim 2$). Ambiguities related to $R_V$ may limit future efforts to measure $H_0$ beyond $\sim 1\%$ precision.  To better resolve extragalactic extinction laws, we directly measure the wavelength-dependent absorption of background galaxies seen in {\it HST} and {\it JWST} images (0.5--2.7\,$\mu$m).
We take the following steps:
\textit{(i)} subtract foreground stars to measure accurate photometry of background galaxies with a tool, \sphot;
\textit{(ii)} measure their redshifts and spectroscopic features with Keck/DEIMOS;
\textit{(iii)} determine their intrinsic spectral energy distributions from the empirical templates which match the absorption lines and breaks in observed spectroscopic features, and
\textit{(iv)} measure $R_V$ by fitting the extinction model to the difference between the template and the observed SEDs.
The above steps are tested with artificial datasets to ensure they accurately recover the input $R_V$.
We apply this set of steps to a first case, NGC 5584, a SN~Ia host and a calibrator of the Hubble constant.
The estimated value of $R_V$ for NGC~5584, $R_V=3.59^{+0.99}_{-0.62}(\text{stat})\pm0.19(\text{syst})$, is consistent with the MW-like extinction law, and it is $\gtrsim 3.5\sigma$ away from $R_V=2$ as favored by SN Ia. If additional hosts show similar results, it would suggest that SN~Ia extinction may not be solely due to mean interstellar dust.  We are now undertaking a statistical study of 5-10 SH0ES hosts to determine the distribution of host extinction laws.
\end{abstract}


\section{Introduction} \label{sec:intro}
    The Hubble constant (H$_0$) is a measurement of the current expansion rate of the universe, and it anchors the expansion history of the universe in cosmology. 
    The \texttt{SH0ES} (Supernovae and H$_0$ for the Equation of State of dark energy) measurements of the local Hubble constant \citep{Riess_2022_SH0ESmain, Murakami_2023_SIP, Breuval_2024_SMC} uses the luminosity distance to Type Ia supernovae \citep[SNe~Ia; see, e.g., ][]{Filippenko_2005_SNIacosmology, Branch_Wheeler_2017_Supernova} and their host galaxy's redshifts to derive H$_0$.
    The measurement of the luminosity distance relies on an accurate calibration of the absolute magnitude of the SN~Ia, and this is achieved by the distance ladder that cross-calibrates SNe~Ia with other nearby distance indicators, such as Cepheid variable stars. 
    For this calibration to be accurate, it is essential that (i) the absorption and scattering of light --- extinction --- by dust grains in the interstellar medium (ISM) of host galaxies is corrected, and (ii) the extinction-corrected luminosity of Cepheids does not change between rungs (steps of calibration). 

    Observations of stars with known brightnesses and colors can reveal the wavelength-dependent profile of the dimming of stars. The observed deviation from the intrinsic brightness measures the dust extinction in the magnitude scale (\textit{total extinction}),
    \begin{equation}
        A_\lambda = m_{\lambda,\text{obs}} - m_{\lambda,0}
        = -2.5\log\left(\frac{F_{\lambda,\text{obs}}}{F_{\lambda,0}}\right)\, .
    \end{equation}
    The profile of $A_\lambda$ depends on the column density of dust and the wavelength-dependent property of dust itself. It is therefore useful to isolate the wavelength dependency of the extinction by normalizing at the $V$ band,
    \begin{equation} \label{eq:extinction_law}
        \xi_V(\lambda) = \frac{A_\lambda}{A_V}\, .
    \end{equation}
    This profile is called the \textit{dust extinction law} \citep[for reviews, see][]{Savage_Mathis_1979_ISMreview, Draine_2003_dust_review, Galliano_2018_dust_review, Salim_2020_dustreview, Gordon_2023_dustlaw}. At optical wavelengths, the profile approximately follows $\xi\propto \lambda^{-1}$, which suggests that smaller grains of dust are more abundant compared with larger grains. \cite{Mathis_1989_dustlawRV} found that various empirical dust extinction laws can be characterized by a single parameter, the slope of the optical extinction law 
    \begin{equation} \label{eq:RV_definition}
        R_V = \frac{A_V}{A_B - A_V} = \left(\frac{A_B}{A_V}-1\right)^{-1}\, .
    \end{equation}
    Various functional forms of the dust extinction law $\xi_V(\lambda,RV)$ that only depend on $R_V$ have been proposed by many extensive studies over three decades \citep{ODonnell_1994_dustlaw, Calzetti_2000_dustlaw, Fitzpatrick_1999, Fitzpatrick_2004_review, Fitzpatrick_Massa_2007_dustlaw, Gordon_2023_dustlaw}.
    The role of $R_V$ is to describe the property of dust and is deeply tied to the grain-size distribution --- a smaller value of $R_V$ corresponds to a stronger preferential extinction toward shorter wavelengths (i.e., grain-size distribution skewed toward small scales), and a larger value of $R_V$ corresponds to a flatter, grayer extinction (i.e., more uniform grain-size distribution). 
    
    In many cases, astronomical observations \textit{require} the correction of dust extinction to avoid unwanted biases in the luminosity and color measurements. 
    The correction is done by estimating $A_\lambda$ so that the intrinsic brightness can be calculated as $m_{\lambda,0} = m_{\lambda,\text{obs}} - A_\lambda$. Equations~\ref{eq:extinction_law} and \ref{eq:RV_definition} give
    \begin{equation}
        A_\lambda = R_V \cdot \left(A_B - A_V\right) \cdot \xi(\lambda,R_V)\, ,
    \end{equation}
    and this total extinction depends on two quantities, $R_V$ and $(A_B-A_V)$.
    The quantity $(A_B-A_V)$ can be estimated observationally since it is equivalent to the deviation of observed color from the known color of the similar object (\textit{color excess}),
    \begin{align}\nonumber
        E(B-V) & \equiv A_B - A_V \\ \nonumber
               & = (m_{B,\text{obs}} - m_{B,0}) - (m_{V,\text{obs}} - m_{V,0})\\ 
               & = (m_{B,\text{obs}} - m_{V,\text{obs}}) - (m_{B,0} - m_{V,0}) \ .
    \end{align}

    The slope of the dust extinction law in the MW on average is found to be $R_V \approx 3.1$--3.3 using observations of O- and B-type stars \cite[][]{Savage_Mathis_1979_ISMreview,Fitzpatrick_1999,Schlafly_2016_MWdust}.
    
    In modern observational cosmology, however, the bright sources used to measure the distance to galaxies (``distance indicators''), such as Type Ia supernovae (SNe~Ia) and Cepheid variable stars in galaxies outside the MW, go through an additional layer of dust within their host galaxies before reaching the MW \citep[e.g.,][]{Brout_2022_PantheonPlus}. 
    The dust extinction law depends on the dust grain distribution and their properties as discussed previously, and host galaxies with ISM-related properties (e.g., star-formation rate; SFR) different from the MW may have a dust extinction law other than $R_V \approx 3$. 

    Measurements of nearby (dwarf) galaxies indeed confirm that the dust extinction law could be different from that of the MW. Reported values of extinction laws in the Large Magellanic Cloud (LMC) and Small Magellanic Cloud (SMC), such as $R_V=3.4$ \citep[LMC average;][hereafter G03]{Gordon_2003_dustlaws_MW_SMC_LMC},  $R_V=2.76$ (LMC2 supershell; G03, c.f. \citealt{DeMarchi_2016_LMCdustlaw} who claim $R_V=4.5$), $R_V \approx 2.7$ \citep[SMC bar;][G03]{Bouchet_1985_SMC_extinction}, $R_V=2.05$ (SMC wing; G03) show that the extinction law can vary depending on the local environment\footnote{We focus on the slope in this work considering the shortest wavelength in our dataset is the $V$ band, but it is worth noting that there is another significant variation in the ``bump'' feature near 2175\,\AA, in addition to the slope, which is present in the MW and LMC but not strongly in the SMC extinction law.} \citep[see further measurements and discussions given by][]{Fitzpatrick_1986_LMC, YanchulovaMJ_2017_SMCextinction, Wang_2023_SMCLMC_extinction_HST}.
    Despite this diversity, 
    there are proxies for the dust extinction law
    --- extragalactic dust extinction laws appear to be correlated with the SFR and the stellar mass formed in each galaxy \citep{Salim_2018_GSWLC2,Hahn_2022_EDAdust}\footnote{We note that these studies are primarily focused on the dust \textit{attenuation} law, which is deeply tied to and similar to the extinction but includes additional light back-scattering into the observed line of sight from nearby sources. In cosmology, we solely focus on the \textit{extinction} law, since the brightness of transient objects, such as SNe~Ia, is measured by subtraction from a template image and this additional effect of back-scattering is canceled.}.  
    
    This galaxy vs. extinction relation allows one to customize the extinction law based on the galaxy types (measured by, for example, morphology, stellar mass, and SFR) to estimate the $R_V$ value for each galaxy. 
    The \texttt{SH0ES} distance ladder therefore strictly selects hosts that are MW-like, spiral, star-forming galaxies for Cepheids and SNe~Ia. Thus, the extinction laws in hosts are expected to be MW-like (i.e., $R_V \approx 3.3$)\footnote{Note that the progenitors of Cepheids are O- and B-type stars, meaning that Cepheids belong to the same stellar population from which $R_V$ values are derived in the MW.}, and its near-infrared (NIR) equivalent value $R_H \approx 0.36$ \citep[see][for definition]{Riess_2022_SH0ESmain} is used to measure the reddening-free magnitude of Cepheids \citep[for a review, see][]{Madore_1982_Wesenheit}, which are then used to calibrate SNe~Ia. 
    
    Similarly to Cepheids, SNe~Ia also require a reddening correction \citep[e.g.,][]{Riess_1996_MLCSs}. This process, as part of the standardization to correct differences between individual SNe, employs a single, $R_V$-like parameter \citep[``color coefficients": e.g., $\beta$ for SALT; ][]{Guy_2007_SALT2, Kenworthy_2021_SALT3}.
    A series of discussions and analyses of the post-standardization residuals and their host environments \citep{childress_2013_massstep, Rigault_2020_SFRdependency, 
    Murakami_2021_SNIa_host, Zhang_2021_ageslope, Brout_Scolnic_2021, Popovic_2021_hostBBC} have further shown that this reddening may be due to the ISM dust in hosts. 
    Interestingly, the color coefficients (and subsequent estimates of $R_V$) from SN~Ia statistics point to $R_V\lesssim2$ ($R_V \approx 1.75$, \citealt{Nobali_Goobar_2008_SNcolor_RV}; $R_V<2$, \citealt{Burns_2018_CSP}; $R_V \approx 2.1$, \citealt{Smadia_2024_SNIa_RV}), a steeper extinction law than the MW average.

    \begin{figure*}
        \centering
        \includegraphics[width=\linewidth]{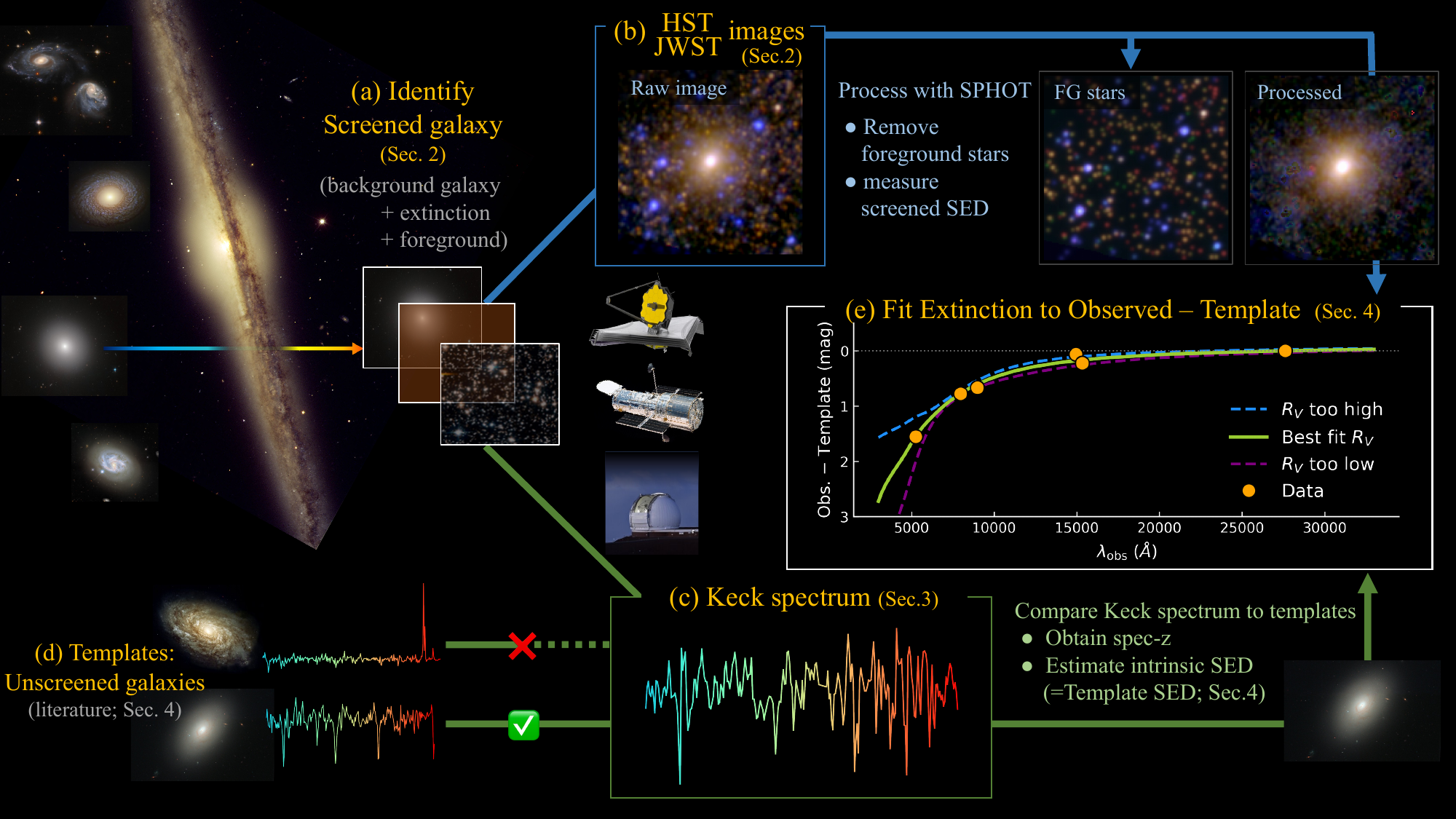}
        \caption{A conceptual illustration and an overview of this paper. The workflow for photometry (blue, top) and spectroscopy (green, bottom) flows from left to right. The photometry measures the observed SED, and the spectroscopy (aided by literature templates) estimates the intrinsic (``template") SED. The extinction law is fitted to the measured difference between the observed and template SEDs.}
        \label{fig:concept}
    \end{figure*}
    
    The lower $R_V$ value in SN analyses alone is not necessarily surprising --- SNe~Ia, likely originating from binary systems involving one or more white dwarfs, are significantly older ($\sim 10^0$--$10^1$ Gyr) than Cepheid variables ($\sim 10^1$--$10^2$ Myr); consequently, SNe~Ia are found farther from the disk at greater scale heights. These differences are enough to make the local ISM environments around them different. Yet there is no evidence that such a significant difference in $R_V$ exists within a galaxy on average, and a recent study by \cite{RinoSilvestre_2025_RVpolarimetry} suggests the MW-like $R_V$ at SN~Ia locations\footnote{Note that the authors measured the extinction law from polarimetry, which is only applicable to absorption-dominated extinction. Care must be taken to extrapolate the result to the generic extinction that includes scattering, and further studies are eagerly awaited.}.
    Thus, identifying the physical origin of the difference (e.g., circumstellar medium, local metallicity) and confirming the consistency of $R_V$ for Cepheids has a significant impact on cosmological studies.

    The absolute distance scale to Cepheids are determined by four geometric anchors (MW, LMC, SMC, and NGC~4258) so that the distance to the SN calibrators (SN hosts with Cepheids) can be measured.
    If for some reason the mean extinction law of calibrators is different from the mean extinction law of geometric anchors, this needs to be accounted for in order to maintain the consistency between rungs in the distance ladder. For example, treating the Cepheids in SN calibrators with a lower mean value of $R_V \approx 1.5$ while keeping the anchors at $R_V \approx 3$ could lower the value of $H_0$ by $\sim 1$~km~s~Mpc$^{-1}$ \citep{Mortsell_2022_H0revisited, Mortsell_2022_H0dust}. In addition to the effect of mean $R_V$, if a wide range of $R_V$ is found within SN calibrators, an additional systematic uncertainty needs to be included in the final H$_0$ estimates. Alternatively, if the $R_V$ of SN calibrators is found within a narrow range near the MW-like extinction ($R_V \approx 3$), this eliminates a fraction of the systematic uncertainty included in the SN~Ia analysis, further tightening the H$_0$ measurement.
    
    Measuring the extinction law requires sources with identifiable intrinsic color (spectral energy distribution, SED).
    Some existing methods make use of quasars in the background \citep{Ostman_2008_quasar_extinction,Menard_2010_quasardust} or visually overlapping galaxies \citep{Holwerda_2017_occult}, but these methods are limited by the availability of such sources and are not suited to measure the extinction law of a particular galaxy.
    We aim to overcome the challenges and measure the dust extinction law of a Cepheid-SN~Ia host by using background galaxies of measured redshifts as the SED source. Background galaxies, seen in \textit{HST} and \textit{JWST} images through the foreground galaxy's disk, are abundant \citep{Holwerda_2005_galaxy_counts} and independent from the stellar population of the foreground galaxy. The extinction by the foreground disk makes them appear redder and fainter than they are expected to be at their spectroscopic redshifts, and this allows us to measure the extinction law.
    The {\it JWST} programs (GO-1685, GO-1995, GO-2875) to observe SH0ES galaxies \citep{Riess_2022_SH0ESmain, Riess_2024_JWSTcrowding} in the NIR provide an unprecedented opportunity to conduct such measurements; the NIR frames at a few microns reveal the nearly unattenuated view of the background galaxies, allowing us to identify them, measure their brightness in {\it HST} optical images, and conduct follow-up spectroscopy to obtain their redshifts.

    The measurement requires four steps (see Figure~\ref{fig:concept}) as follows. 
    (i) Identify and accurately measure the galaxy SEDs from the \textit{HST} and \textit{JWST} images (Fig.~\ref{fig:concept}, panels $a$ and $b$). During this process, the flux from background galaxies is isolated from the foreground stellar light. We present our photometry, newly developed data processing tool, and the test results of its performance in Section~\ref{sec:photometry}. 
    (ii) Conduct targeted, multi-object spectroscopy of the background galaxies with Keck/DEIMOS\footnote{We have also used Keck/LRIS in subsequent observing runs.} (Fig.~\ref{fig:concept}, panel $c$). We measure the redshift of the background galaxies and prepare continuum-removed spectra for the spectroscopic comparison in the next step. The spectroscopic data reduction processes are discussed in Section~\ref{sec:spectroscopy}. 
    (iii) Estimate their intrinsic SEDs from the unscreened empirical SED templates  (Fig.~\ref{fig:concept}, panel $d$). 
    The intrinsic SED is estimated from a linear combination of spectrophotometric templates, weighted by the spectral similarity of each template to the observed spectrum.
    This procedure is extensively tested along with the next step below.
    (iv) Fit the dust extinction law for the foreground layer. We optimize the mean $R_V$ value that best describes the dimming of the background galaxies due to the foreground dust, using the functional form of dust extinction profile (Eq.~\ref{eq:extinction_law}) by \cite{Fitzpatrick_1999}. The total extinction $A_V$ values are fitted for individual background galaxies, simultaneously with the $R_V$ value. This method, along with the test results, is presented in Section~\ref{sec:analysis}.
    
    Finally, we apply all these methods on the background galaxies seen through NGC~5584, a SH0ES galaxy at $\sim 22$~Mpc, to calculate the optical dust extinction slope $R_V$; the results are shown in Section~\ref{sec:results}. We present possible extensions and improvements in Section~\ref{sec:discussion}, followed by a conclusion in Section~\ref{sec:conclusion}, where we discuss the implications of the results for NGC 5584 and review the performance of this method for future studies.

\section{HST and JWST photometry} \label{sec:photometry}
    
    \begin{figure*}
        \centering
        \includegraphics[width=\linewidth]{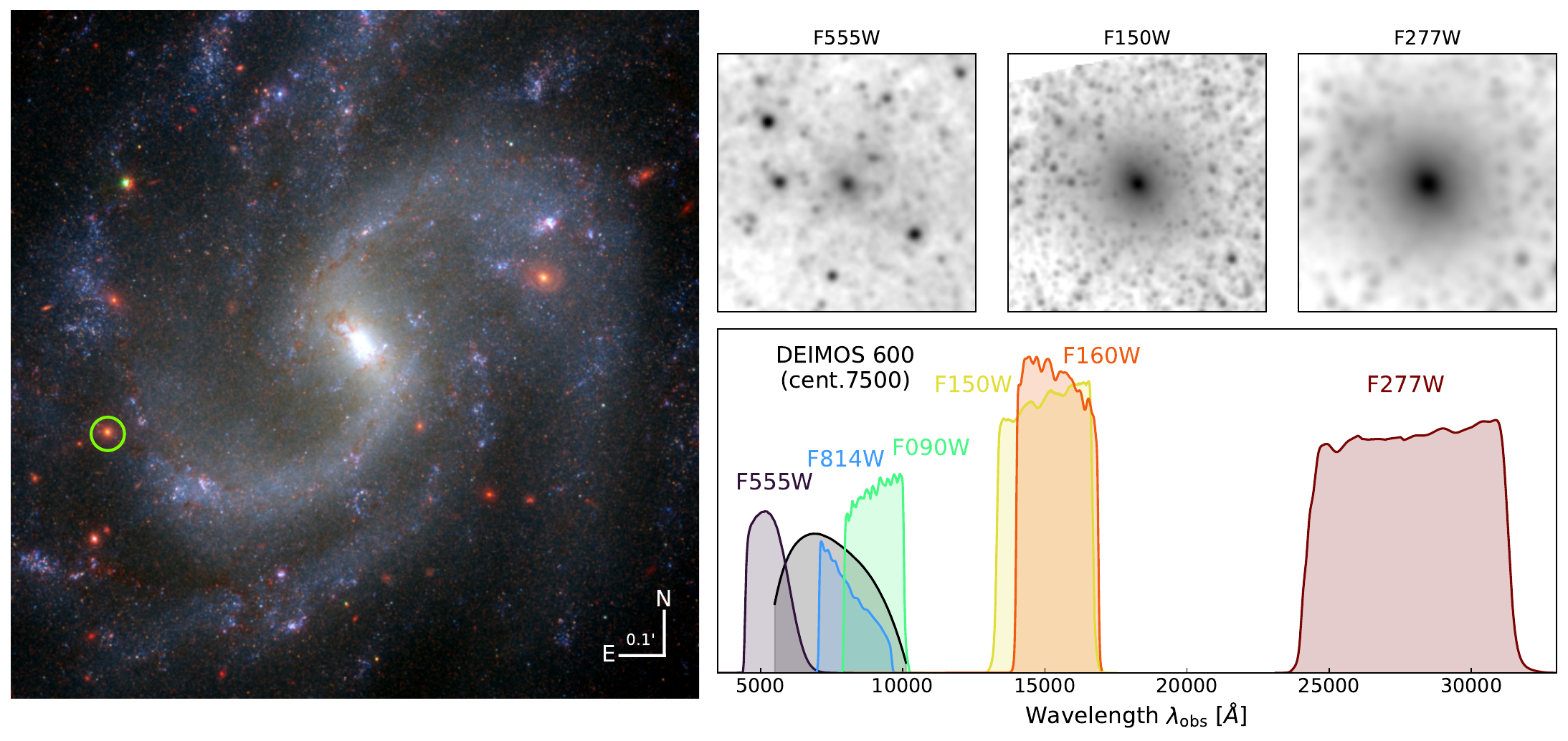}
        \caption{\textit{Left:} A composite image (blue, F555W; green, F814W; red, F277W) of the foreground galaxy NGC 5584. Background galaxies are seen as red extended sources. \textit{Top right:} cutouts of a selected background galaxy (g260, circled) at F150W, F150W, and F277W. \textit{Bottom right:} The transmission functions of \textit{HST} and \textit{JWST} filters used in this study as well as Keck/DEIMOS with the configuration used in our observation.}
        \label{fig:concept_cutouts}
    \end{figure*}
    
    \begin{figure*}
        \centering
        \includegraphics[width=\linewidth]{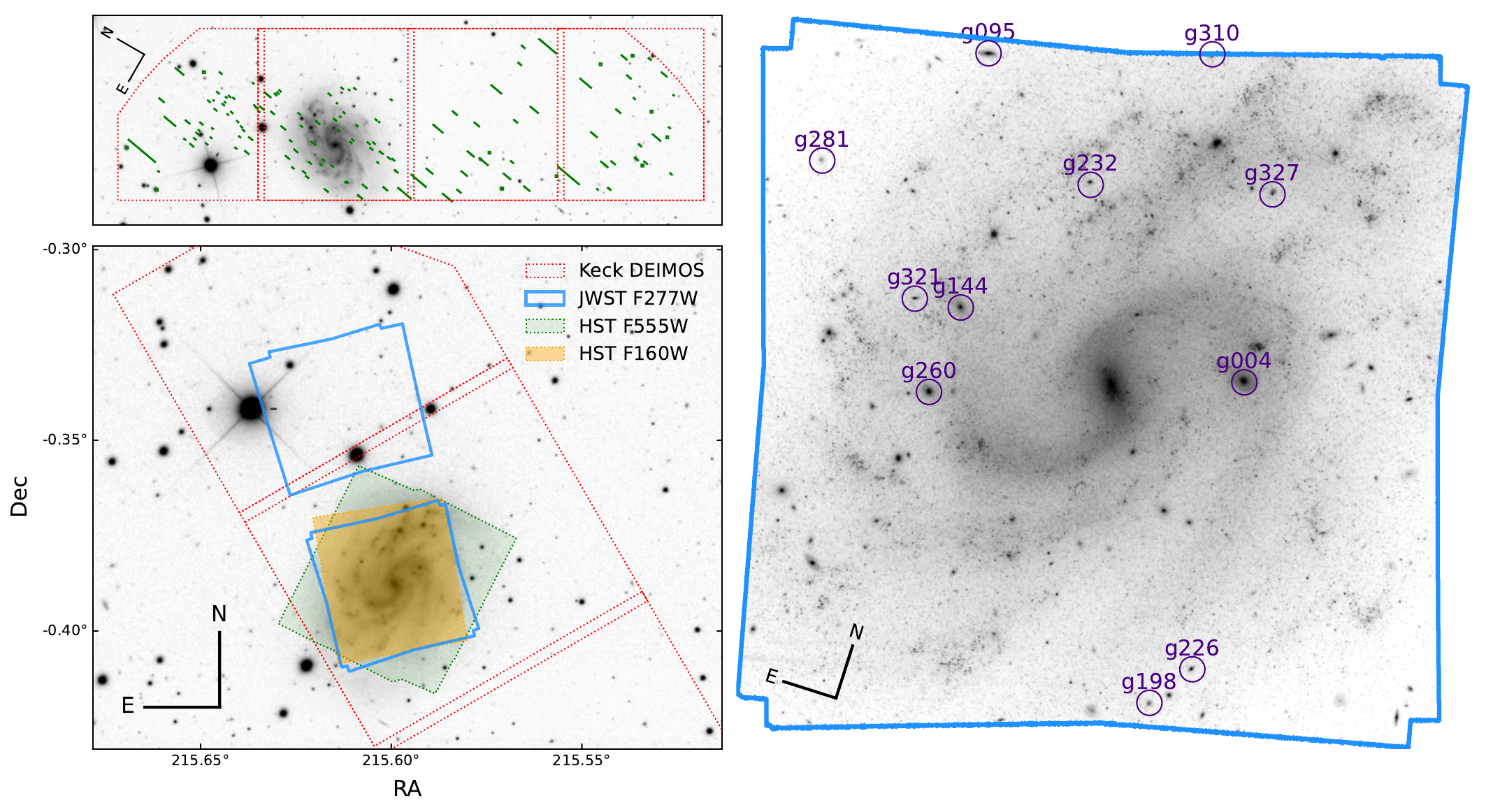}
        \caption{Our observation layout. \textit{Bottom left:} image of the sky around NGC 5584; north is up, east to the left. The approximate field of view for each of the three instruments (Keck/DEIMOS, \textit{HST}/WFC3, and \textit{JWST}/NIRCam) is shown. \textit{Top left:} the slitmask design for Keck/DEIMOS observations. The slits are configured to cover as many background galaxy candidates as possible, and the outer regions are used for future studies. \textit{Right:} the background galaxies whose redshift is confirmed by the spectroscopy.}
        \label{fig:observation_layout}
    \end{figure*}

    Our target foreground galaxy, NGC 5584, is a barred spiral located in the Virgo constellation. It is the host galaxy for the Type Ia SN~2007af \citep{Nakano_Itagaki_2007_SN2007af, Salgado_2007_SNclassification}, which is one of the 42 SNe~Ia used in the SH0ES distance ladder \citep{Riess_2022_SH0ESmain} to calibrate their absolute luminosity using Cepheid variables. 
    Recent observations of NGC 5584 with a \textit{JWST}/NIRCam program (GO-1685, PI A. G. Riess; \citealt{Riess_2021_jwst_cycle1}) have complemented the existing optical-to-NIR observations by \textit{HST}/WFC3, enabling a clear identification of background galaxies seen through NGC 5584 and providing a much longer wavelength baseline to measure the background galaxies' SEDs (Fig.~\ref{fig:concept_cutouts}). 
    Designed to be a part of the SH0ES distance ladder, the images of NGC 5584 are optimized for Cepheid variable stars and other distance indicators \citep[e.g.,][]{Li_2024_JAGB,Ganandeep_2024_TRGB_JWST}. 
    This dataset together contains six broadband filters spanning from optical to NIR (F555W, F814W, F090W, F150W, F160W, and F277W), each taken over multiple epochs. Figure~\ref{fig:observation_layout} shows the footprint of the space telescope images, as well as the slitmask design for ground-based spectroscopic observations which we describe in Section~\ref{sec:spectroscopy}.

    We select the background galaxies with the
    following criteria: (i) the object is seen through the foreground galaxy NGC 5584 in all filters, (ii) it has a clear, extended appearance, and (iii) it is isolated from star clusters or star-forming regions. 
    The final selection of the background galaxy candidates is based on the spectroscopic identification of redshifted galaxy features (see Sec.~\ref{sec:spectroscopy}) and is shown on the right panel of Figure~\ref{fig:observation_layout}.

    We measure the SED of the target galaxies using aperture photometry \citep[see, e.g.,][for practices of aperture photometry for galaxies and applications in large surveys]{Strauss_2002_SDSS_MGS}. Since foreground light from stars in the nearby galaxy is superimposed on all of the selected background galaxies,
    removing the foreground light from our target field is essential so that the majority of the remaining flux is from the background galaxies. 
    
    \begin{figure*}
        \centering
        \includegraphics[width=\linewidth]{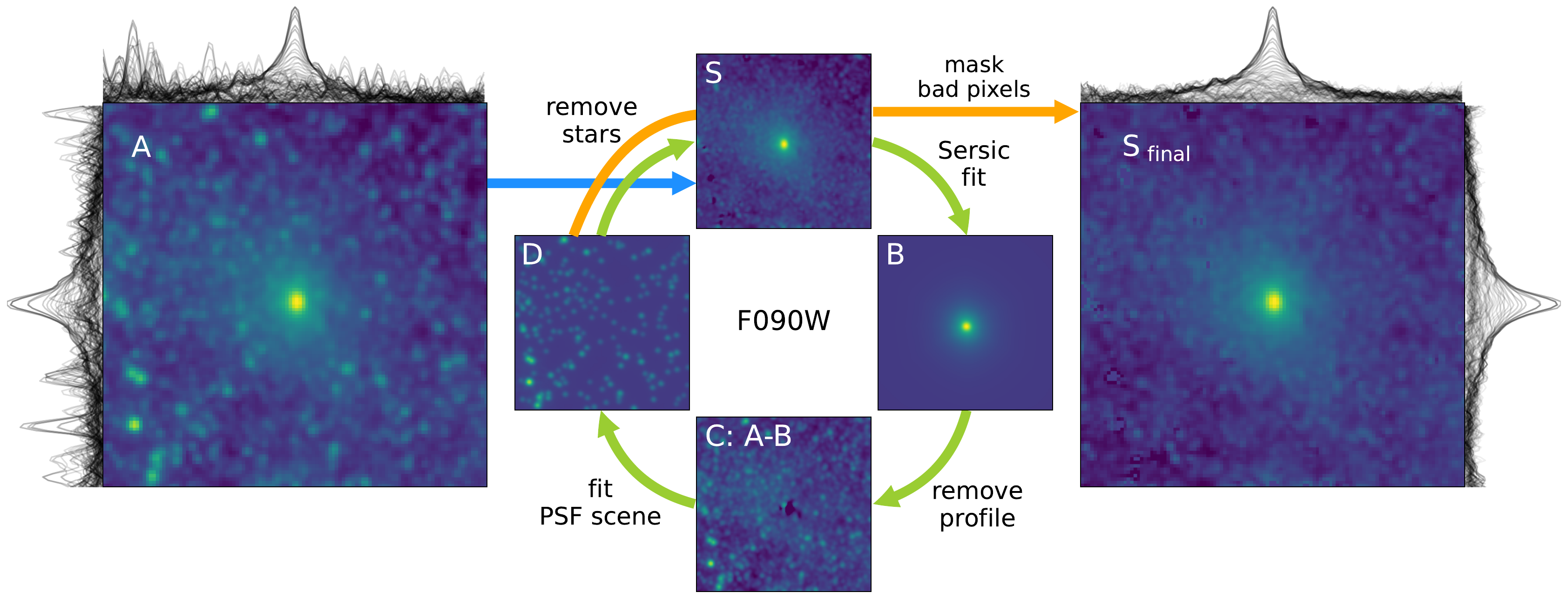}
        \caption{Workflow of \sphot, our custom photometry tool. {\it Left:} The raw data of the target galaxy. One-dimensional slices of the flux, in log scale, are shown along with each axis of the image. The target galaxy is located at the center, but the foreground stellar field takes a larger fraction of the flux, making the photometry difficult. {\it Middle panels:} The iterative process of fitting the \sersic profile (\texttt{B}), subtracting \sersic profile (\texttt{C}), fitting foreground stellar PSF scene (\texttt{D}), and  subtracting the stellar PSFs from the data (\texttt{S}). In each iteration only the PSF and \sersic models are subtracted from the \textit{raw data}, so that fit results are not affected by previous iteration. {\it Right:} The final result of the foreground cleaning. The target galaxy is now clearly visible, and the surrounding ``background'' regions exhibit significantly smaller noise compared with the raw data. Photometry is performed on this final product. False-color images of our target galaxies before/after \sphot processing are shown in Figure~\ref{fig:RGB_cutouts}.}
        \label{fig:sphot-loop}
    \end{figure*}

    \begin{table}
        \centering
        {\tiny $\ $\\}SPHOT Algorithm
        \begin{tabular}{lll}
        \hline
        \hline
        & Description & Definition \\
        \hline
        \multicolumn{3}{l}{Initial, temporary processing}\\
        1.&Raw cutout & \texttt{A = image (N,N)} \\
        2.&Estimate sky & \texttt{F = <Annulus(A)>}\\
        3.&Subtract sky & \texttt{S = A - F} \\
        \hline
        \multicolumn{3}{l}{Iteratively improve \texttt{S} (``galaxy-only" image)}\\
        4.&Fit \sersic model$^\alpha$ & \texttt{B  = Fit(S; PSF $\circ$ S\'ersic)}\\
        5.&``Star-only" image & \texttt{C  = A - B - F}\\
        6.&Fit stellar PSFs & \texttt{D  = Fit(C; $\sum$PSF)}\\
        7.&``Sky-only" image & \texttt{E  = A - B - D}\\
        8.&Residual mask$^\beta$ & \texttt{M  = Bool(|E| < 4$\sigma_E$)}\\
        9.&Update sky model & \texttt{F  = Fit(Ring $\circ$ ME; Poly)}\\
        10.&Update main image  & \texttt{S = M(A - D) - F}\\
        \hline
        11.&Final, science image & \texttt{S$_\mathrm{\texttt{FINAL}}$}\\
        \hline
        \hline
        \end{tabular}\vspace{1mm}\\
        \begin{scriptsize}
        $\alpha$: Full parameters fit for the base filter, flux scale fit for other filters.
        $\beta$: Pixels within $1-R_e$ isophot region are not masked.
        \end{scriptsize} 
        \caption{An overview of SPHOT images and algorithm. The steps 4--10 are repeated to iteratively improve \sersic, stellar PSF, and sky model fits. \texttt{Fit(W;X)} represents optimizing parameters for model \texttt{X} to fit the image \texttt{W}. Convolution of a kernel $Y$ onto image $Z$ is denoted as \texttt{Y$\circ$Z}. \texttt{Bool(C)} is a Boolean image in which each pixel has value of 1 if \texttt{C} is true, otherwise 0. Descriptions of each step can be found in Appendix~\ref{appendix:photometry}.}
        \label{tab:sphot_algorithm}
    \end{table}

    \subsection{SPHOT -- Accurate Photometry of Galaxies Seen through Foreground Stellar Field} \label{sec:sphot}
    We develop a custom photometry tool, \sphot\footnote{https://github.com/SterlingYM/sphot}; see Appendix~\ref{appendix:photometry} for details. The purpose of this tool is to fit the galaxy profile and the stellar point-spread function (PSF) to the data, and to iteratively improve both to eventually reach the best possible PSF subtraction to obtain the ``cleaner'' image with less foreground contamination.
    For the galaxy profile, we use a \sersic profile \citep{Sersic_1963,Sersic_1968_atlas} convolved with each filter's PSF \citep{Geda_2022_petrofit}.

    An overview of the algorithm of \sphot is shown in Table~\ref{tab:sphot_algorithm} and Figure~\ref{fig:sphot-loop}. \sphot iteratively improves \sersic model (\texttt{B}) and stellar PSF fits (\texttt{D}) by providing ``galaxy-only'' (\texttt{S}) and ``star-only'' (\texttt{C}) images for fitting (respectively), while simultaneously updating the residual \texttt{E} for masking bad pixels and fitting the sky gradient \texttt{F}. Once converged, the image \texttt{S$_{\texttt{FINAL}}$} is a background-subtracted, sky-gradient-corrected, and PSF-subtracted image. 
    Since the \sersic profile is only used to improve the stellar PSF subtraction, this final product \texttt{S$_{\texttt{FINAL}}$} retains the morphology of the galaxy and is not significantly dependent on the model.    
    
    \subsection{Aperture Photometry} \label{sec:aperture_photometry_maintext}

    \begin{figure}
        \centering
        \includegraphics[width=\linewidth]{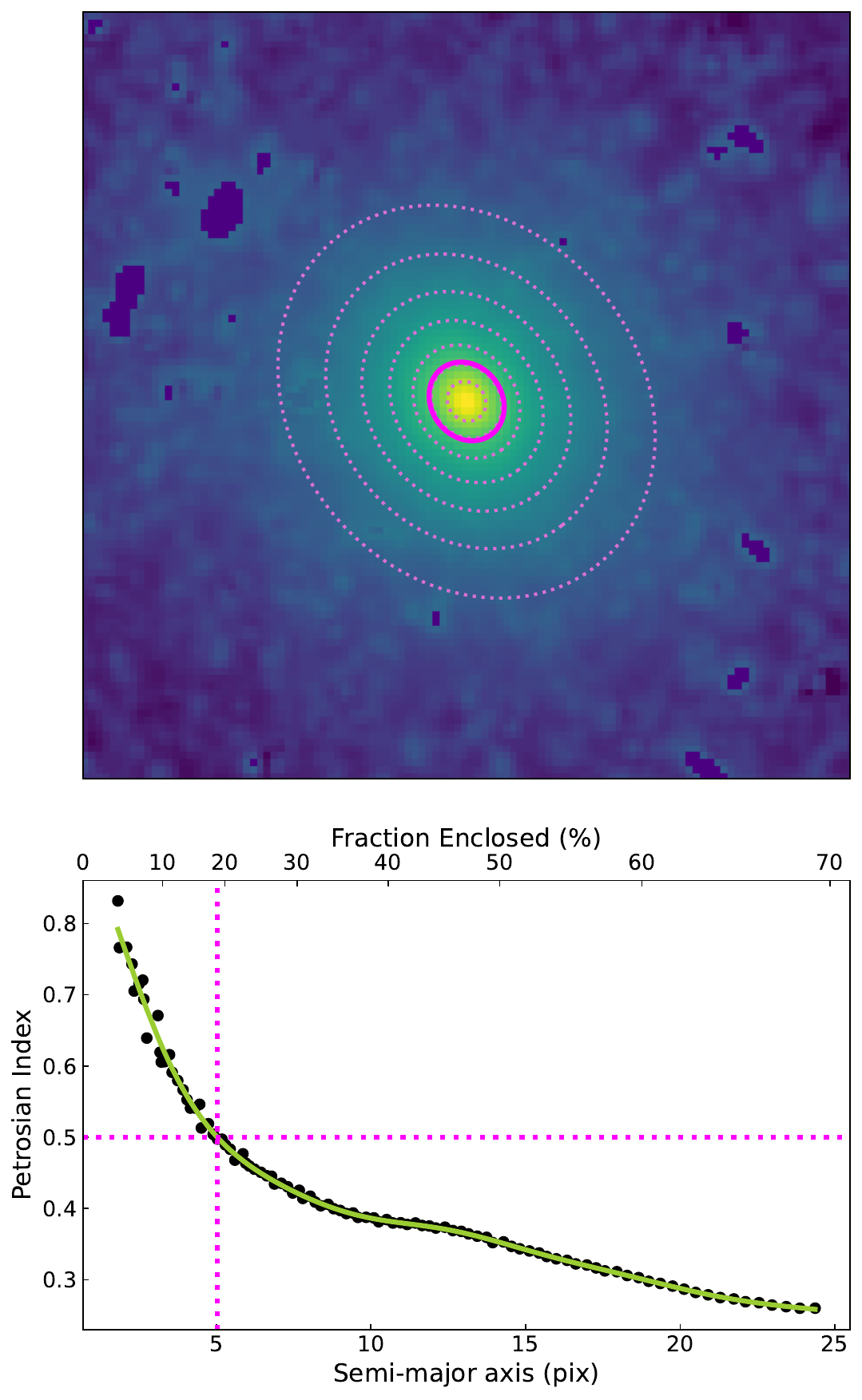}
        \caption{The procedure of determining the aperture size with Petrosian indices. \textit{Top:} the \sphot-processed image of our target galaxy \texttt{g260}. Dotted ellipses represent the isophotal apertures at the enclosed flux levels of 10\%, 20\%, $\cdots$, 70\% of the Petrosian radius (same as the top ticks in the bottom panel). Flagged pixels are marked as solid-fill purple. \textit{Bottom:} the Petrosian index as a function of the aperture size. The measured Petrosian indices are then used to determine the aperture size that corresponds to the desired Petrosian index ($0.5$ in this example).}
        \label{fig:aperture_photo}
    \end{figure}

    Once the images are processed, we perform aperture photometry on the science image \texttt{S}.
    Owing to the nature of the crowded foreground and extinction, the average signal-to-noise ratio (S/N) of the target galaxies is lower than deep universe images of the same depth even after processing with \sphot. This means that a larger aperture is not desired. Unlike many galaxy studies, where the goal of photometry is to measure the total flux from the galaxy \citep[e.g.,][]{Strauss_2002_SDSS_MGS,Barro_2013_CANDELS}, our study only requires the color of the galaxy --- that is, the ratio of flux between filters within the same aperture. We therefore use a smaller aperture with the aperture size determined by the Petrosian indices \citep{Petrosian_1976,Graham_2005_petrosianmag}. 
    As shown in Figure~\ref{fig:aperture_photo}, we calculate the Petrosian indices along the semimajor axis of elliptical isophotes \citep{Jedrzejewski_1987_isophotes} in the F150W image. We test various aperture sizes near the Petrosian indices of 0.3--0.6 and use $0.5$ as the baseline result (discussed  later in Sec.~\ref{sec:discussion} and Figure~\ref{fig:petro_RV_endtoend}). This range corresponds to $\sim 10$--70\% of the total flux, depending on the profile and morphology. For our target galaxies, the baseline aperture encloses approximately a third of the total flux (see Table~\ref{tab:target_galaxies}). The results from the photometry with each Petrosian index are compared and discussed in Section~\ref{sec:discussion}. We use \texttt{Photutils} \citep{Bradley_2022_photutils} to calculate the flux and uncertainty, which includes the photon count noise from the raw data ($\sigma_\text{count}$), the fitting uncertainty from \sphot ($\sigma_\texttt{sphot}$), and the sky variance measured by placing same-sized apertures in the sky region ($\sigma_\text{skyvar}$). Those uncertainties are added in quadrature,
    \begin{equation}\label{eq:aperphot_uncertainty}
        \sigma_{m,\text{aper}}^2 = \sigma_{\text{count}}^2 + \sigma_\text{sphot}^2 + \sigma_\text{skyvar}^2\, .
    \end{equation}
    We find that the sky variance $\sigma_\text{skyvar}$ dominates the uncertainty ($\sim 0.01$--0.1 mag, depending on the aperture size and the band). This is expected as the ``sky'' region contains blended or faint foreground stars that cannot be subtracted by \sphot.

    \subsection{Aperture--PSF Correction} \label{sec:aperture_corr}
    The PSFs of our data greatly vary, with their full width at half-maximum intensity (FWHM) spanning $\sim 0.03''$ to $\sim 0.15''$. 
    To correct for the PSF losses from the fixed aperture, we successively convolve the F090W image (which has the sharpest PSF) to the resolution of every other band using a PSF kernel constructed from the individual PSFs. We then perform aperture photometry on these convolved versions of the F090W image to compute the correction for flux losses in the other bands (similarly to \citealt{Leung_2023_NGDEEP1_PSFref}),
    \begin{equation} \label{eq:psf_correction}
        \Delta m_{\text{aper},i} = m^\text{F090W}_\text{original} - m^\text{F090W}_{\text{convolved},i}\, .
    \end{equation}
    The effect of the aperture correction is more significant at a smaller aperture, where the surface brightness profile of the galaxy is steeper. 

    \subsection{Artificial Galaxy Test and the Final Product}\label{sec:phot_testing}

    \begin{figure}
        \centering
        \includegraphics[width=\linewidth]{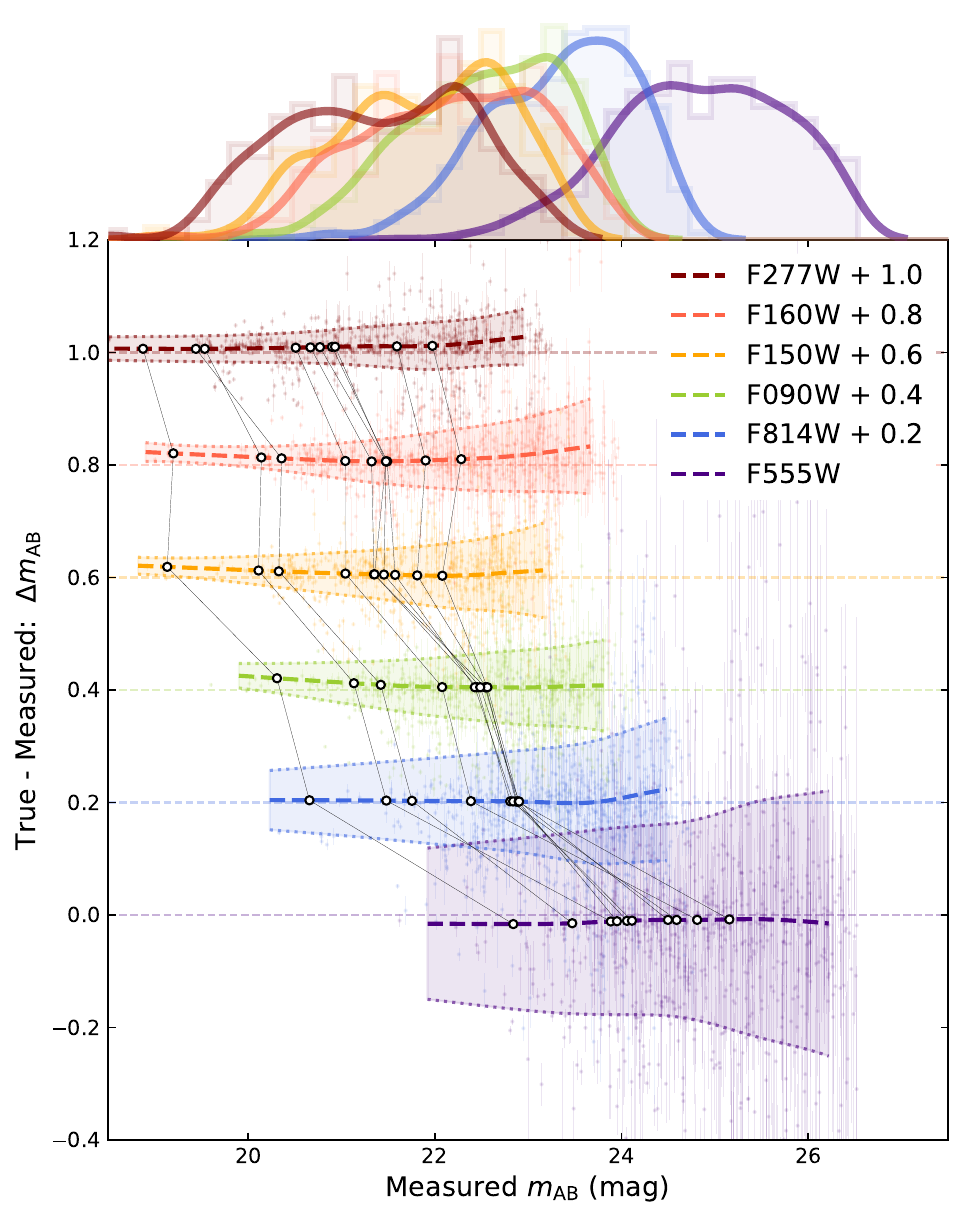}
        \caption{The artificial galaxy test results. The vertical axis represents the deviation of the measured photometry from the true value (see Fig.~\ref{fig:photometry_test_f277w}). The histogram at the top shows the distribution of the test samples in each filter. The colored curves and the shaded regions represent the fitted mean (Eq.~\ref{eq:phot_test_mean}) and the standard deviation of the offsets. Each set of black circles connected by lines indicates the measured brightness of our NGC~5584 background galaxies.}
        \label{fig:phot_bias_all}
    \end{figure}
    
    We evaluate the performance of this workflow with \sphot and aperture photometry by performing an artificial galaxy test. For this test, we superimpose images of galaxies at redshifts similar to our targets to the disk of NGC~5584 and perform the photometry on them.
    Unlike artificial star tests used to measure the effect of crowding \citep[e.g.,][]{Yuan_2020_CephDist_crowding} on point sources, there is no fixed size, profile, or features for background galaxies, and the only way to perform realistic tests is to use an empirical dataset of galaxies at similar redshift ranges taken with similar instruments.
    We therefore use cutouts of deep field galaxies from the \texttt{CANDELS/CEERS} survey \citep{Barro_2013_CANDELS,Finkelstein_2023_CEERS_keyI}.
    Results of the artificial galaxy test are shown in Figure~\ref{fig:phot_bias_all} (see also Fig.~\ref{fig:photometry_test_f277w}). From the test results we derive the correction for the crowding effect, additional uncertainty, and covariance between filters, details of which are described in Appendix~\ref{appendix:artificial_galaxy_test}.
    
    With the corrections (Eq.~\ref{eq:psf_correction} and Eq.~\ref{eq:phot_test_mean}) added, we obtain the measured brightness\footnote{Throughout this paper we use AB magnitudes \citep{Oke_Gunn_1983}: $m_\text{AB}=-6.10 - 2.5\log_{10}(f_\nu/\text{MJy})$. With all of the images reprojected and pixel-aligned into $0.03''$/pix scale (for consistency with the \texttt{CEERS} DR0.5 and DR0.6: see Appendix~\ref{appendix:artificial_galaxy_test} for discussion), this corresponds to the zeropoint of 28.08652.} 
    of our target galaxies across all bands $m^\text{obs}_\text{corr}$ (Eq.~\ref{eq:mag_obs}) and their covariance matrix $C_\text{phot}$ (Eq.~\ref{eq:phot_cov}); see Appendix~\ref{sec:photometry_correction} for formulae.

    \begin{figure*}
        \centering
        \includegraphics[width=\linewidth]{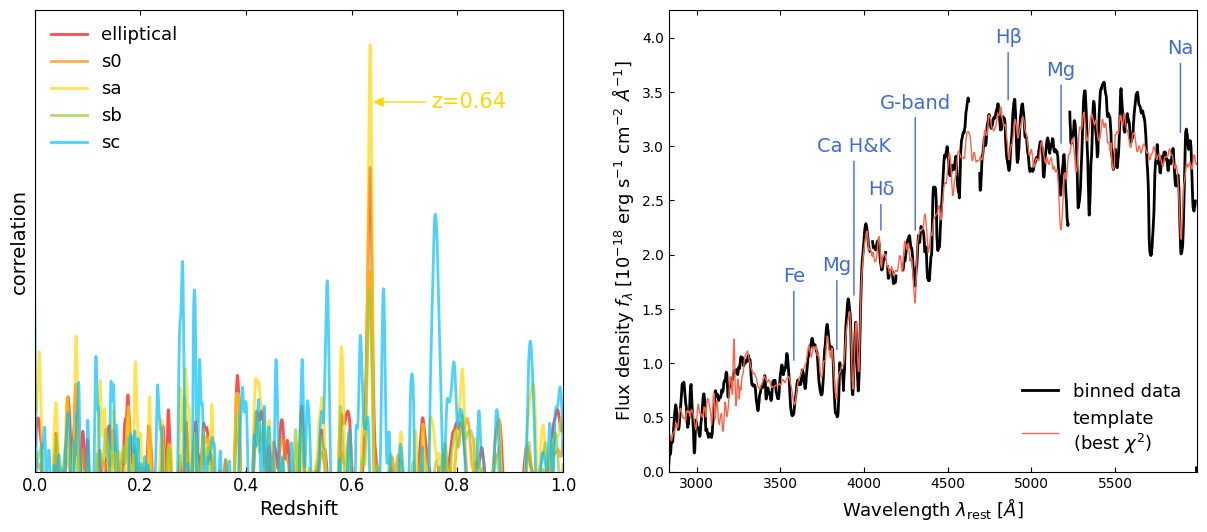}
        \caption{The two-step redshift measurement. 
        \textit{Left:} The redshift and the best-fit template are identified from the peak of the cross-correlation values across the redshift range.
        \textit{Right:} An example of the observed spectrum (g310) and the best-fit template spectrum. The redshift is determined by minimizing the $\chi^2$  value between the observed spectrum and the template. The continuum of the observed flux is added back to aid visibility (e.g., near the 4000\,\AA\ break). }
        \label{fig:spec_obs_cross_z}
    \end{figure*}

\section{Keck/DEIMOS Spectroscopy} \label{sec:spectroscopy}

    We observed NGC 5584 with Keck/DEIMOS \citep{Faber_2003_DEIMOS} using the custom-designed slitmask (Fig.~\ref{fig:observation_layout}, top-left panel) on 2023 May 11 and 12 UTC. Each night we obtained $4 \times 1200$\,s exposures with this slitmask, making the total exposure time $9600$\,s. 
    The slitmask designs and target selections, such as object separations, slit lengths, and apparent brightness, are inspired by the \texttt{DEEP2} survey \citep{Newman_2013_deep2}. Target galaxies identified and chosen in the F277W image are given weights by their apparent sizes, which were then used to optimize the target selection in the slitmask design software \texttt{DSIMULATOR}.
    Fields outside NGC 5584 are used to obtain spectra of bright galaxies for future studies. Images from SDSS DR17 \citep{Abdurrouf_2022_DR17} are used for the selection of such galaxies and alignment stars.

    Similarly to photometry, our major task is to identify spectroscopic features of the background galaxies in the presence of foreground stars and gas. 
    We approached this as part of the reduction process, and any remaining emission or absorption lines identified at NGC~5584's redshift, as well as telluric bands, are masked. 
    The process of data reduction and foreground removal is described in Appendix~\ref{appendix:spectroscopy}.
    
    \subsection{Redshift Identification}

    \begin{deluxetable*}{clcccrrrrrrrrrr}
        \vspace{3mm}
        \tablehead{\# &\colhead{ID} & \colhead{Redshift} & \colhead{RA} & \colhead{Dec}& \colhead{F555W} & \colhead{F814W} & \colhead{F090W} & \colhead{F150W} & \colhead{F160W} & \colhead{F277W}
        & \colhead{$r_a$}& \colhead{$r_b$}& \colhead{$\theta$}& \colhead{$F_\text{frac}$}\\
        && $z$ &(deg)&(deg)&($m_\text{AB}$)&($m_\text{AB}$)&($m_\text{AB}$)&($m_\text{AB}$)&($m_\text{AB}$)&($m_\text{AB}$)
        &($''$)&($''$)&(deg)&}
        \startdata 
        1 & g004 & 0.3617 & 215.59235 & -0.38538 & 23.50 & 21.30 & 20.92 & 19.70 & 19.77 & 19.45 & 0.23 & 0.22 & -29 & 0.24 \\
        2 & g095 & 0.6394 & 215.61065 & -0.37279 & 24.28 & 22.38 & 22.05 & 20.85 & 20.85 & 19.84 & 0.66 & 0.24 &  15 & 0.30 \\
        3 & g144 & 0.6353 & 215.60796 & -0.38615 & 25.42 & 23.15 & 22.83 & 21.68 & 21.60 & 20.92 & 0.11 & 0.10 & -69 & 0.21 \\
        4 & g198 & 0.6385 & 215.59201 & -0.40319 & 24.21 & 23.00 & 22.68 & 21.95 & 22.03 & 21.71 & 0.43 & 0.43 &  21 & 0.57 \\
        5 & g226 & 0.5747 & 215.59037 & -0.40079 & 24.98 & 23.11 & 22.77 & 21.63 & 21.76 & 21.02 & 0.25 & 0.19 &  68 & 0.36 \\
        6 & g232 & 0.6312 & 215.60334 & -0.37782 & 24.49 & 23.13 & 22.74 & 21.85 & 21.77 & 21.16 & 0.29 & 0.20 &  22 & 0.42 \\
        7 & g260 & 0.6271 & 215.60821 & -0.39095 & 24.53 & 22.03 & 21.66 & 20.62 & 20.65 & 20.03 & 0.19 & 0.16 & -51 & 0.21 \\
        8 & g281 & 0.6378 & 215.61737 & -0.38092 & 24.43 & 23.35 & 23.01 & 22.48 & 22.70 & 22.36 & 0.42 & 0.40 & -45 & 0.33 \\
        9 & g310 & 0.6354 & 215.59927 & -0.36923 & 25.10 & 22.66 & 22.37 & 21.35 & 21.32 & 20.77 & 0.11 & 0.09 &  82 & 0.25 \\
        10 & g321 & 0.4578 & 215.61044 & -0.38644 & 24.70 & 23.09 & 22.69 & 21.62 & 21.65 & 21.07 & 0.42 & 0.17 & 20 & 0.48 \\
        \enddata
        \caption{The list of our target galaxies, spectroscopic redshifts, and aperture magnitudes with Petrosian index 0.5. The last four columns ($r_a$, $r_b$, $\theta$, $F_\text{frac}$) describe the semimajor axis, semiminor axis, angle of the elliptical aperture, and an approximation of the fraction of the enclosed flux. Note that the estimation of the total flux has a large uncertainty, and the value of $F_\text{frac}$ may not be accurate.}
        \label{tab:target_galaxies}
    \end{deluxetable*}

    Using the obtained spectra, we measure the redshift of each galaxy through cross-correlation and $\chi^2$ minimization (Fig.~\ref{fig:spec_obs_cross_z}). The continuum of each spectrum is approximated with Gaussian smoothing on the line-masked spectrum and is removed.
    We use \texttt{Specutils} \citep{Earl_2023_specutils} to perform  cross-correlation \citep[e.g.,][]{Tonry_Davis_1979_crosscorr} of our spectra against \cite{Kinney_Calzetti_1996} templates.
    The best-fit redshift is subsequently refined with linear regression of the best-correlated template to the observed spectrum over a fine redshift grid. This reduces the effect of grid resolution in the cross-correlation method and provides a more precise redshift, which is critical to reducing systematic effects in the spectral similarity analysis we discuss in Section~\ref{sec:analysis}.
    Figure~\ref{fig:spec_obs_cross_z} shows an example of our observed spectrum and the best-match template. The match between the best-correlating template's morphological type and the observed galaxy's spectral features is used as an additional check.

\section{Inference of the Foreground Extinction} \label{sec:analysis}
    
    \begin{figure*}
        \centering
        \includegraphics[width=\linewidth]{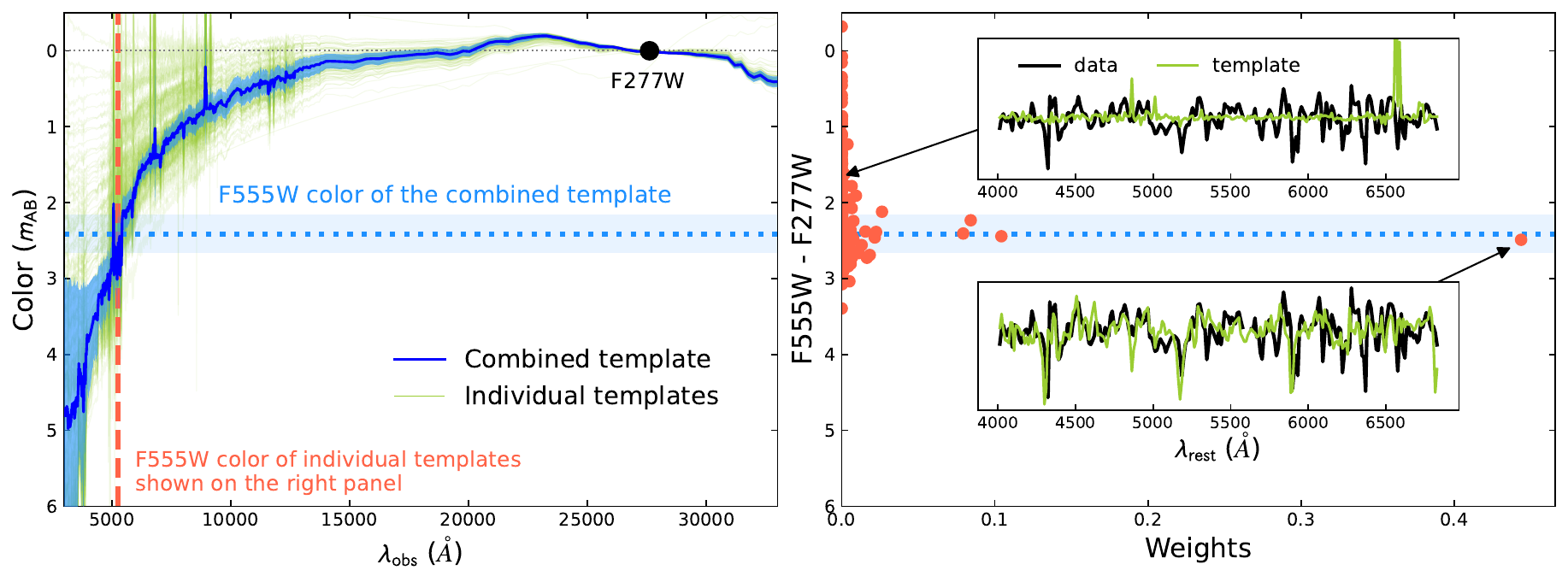}
        \caption{A visual demonstration of the linearly combined template. \textit{Left:} The SED of the combined template in the observed frame, along with individual templates. \textit{Right:} The weights of the templates and their color $F555W-F277W$.}
        \label{fig:weighted_template}
    \end{figure*}
    
    With the photometry and spectroscopy methods described above, we obtain a dataset of background galaxies that are screened by dust in the foreground galaxy, NGC 5584. The list of galaxies, their photometry, and identified redshift are provided in Table~\ref{tab:target_galaxies}. The difference between the observed SED and the intrinsic SED of these galaxies is the dust extinction curve of NGC 5584, and therefore estimating the intrinsic SED plays a critical role in this process. 
    To do this we must model galaxies as complex systems with a wide range of possible stellar populations, internal dust content, and star-formation histories, and the resulting SED varies significantly \citep{Conroy_2013_galaxySEDreview}.

    One enlightening aspect of our dataset is that we have both optical spectra and the photometric SED of the target galaxies. While the photometric SED is what provides the extinction law and therefore cannot be used to estimate the intrinsic SED, the optical spectra can be flattened and used as an SED-free proxy to guide us to the best estimate of the galaxy's intrinsic SED.

    Considering the requirements and the restrictions (see discussion in Sec.~\ref{sec:discussion}), in this work we use an empirical SED template of galaxies to estimate the intrinsic SED of each background galaxy.
    A number of recent studies of galaxy spectra, such as \cite{Portillo_2020_VAE}, \cite{Melchior_2023_SPENDER_I}, \cite{Liang_2023_SPENDER_II}, \cite{Ferreras_2023_spectra_entropy}, or \cite{Martinez_solaeche_2024_eCALIFA}, have shown that spectral features of galaxies provide strong insights into the physical activities, stellar populations, and properties within the galaxy, which largely determine the intrinsic SED of the galaxy. 
    We, therefore, follow the ansatz that \textit{spectroscopically similar galaxies have similar SEDs}, and the intrinsic SED of the target galaxy can be approximated by a linear combination of the template galaxies. The weight of each template is calculated based on the similarity between the template and the observed spectra. For the comparison, we remove the continuum, mask emission lines, and normalize the spectrum as described in the following section.

    \subsection{Linear Combination of Templates} \label{sec:template_combination}
    Empirically modeling the intrinsic SED in this study requires a set of spectral templates of galaxies that allows us to compare them against the observed spectra and corresponding photometric SEDs. The transmission coverage of our photometric SED spans $\sim 3000$--32,000\,\AA, and the optical spectra cover the range of $\sim 4000$--9500\,\AA. The templates therefore need to cover both of these ranges of wavelengths in the observed frame (i.e., redshifted to the target galaxy's redshift $\sim 0.6$). 
    \cite{Brown2014_templates} provide an excellent set of spectrophotometric templates, thanks to their wide range of wavelength coverage and the use of best-fit synthetic models to interpolate observed data. Their careful treatment of matching apertures for spectra and photometry makes this dataset ideal for our study.

    We construct the intrinsic SED of the target galaxy by calculating the linear combination of this template library --- the templates with optical spectra similar to the observed spectra are highly weighted, and templates with unmatching spectral features are deweighted, essentially making the combined SED from spectroscopically similar templates only.
    We describe the methodology and extensive testing in Appendix~\ref{appendix:spectrum-matching}.
    
    \subsection{Extinction-Law Fit}
    With the observed SED ($m_\text{obs}$; Eq.~\ref{eq:mag_obs}) and combined template SED ($m_\text{temp}$; Eq.~\ref{eq:mag_temp}) in hand, we fit the optical extinction parameters ($A_V$ and $R_V$) with a \cite{Fitzpatrick_1999} dust law to each background galaxy.
    
    To account for the overall brightness difference (owing to distance modulus, fraction of enclosed flux, and total stellar mass) between the template and observed SEDs, we include a distance-modulus-like uniform offset,
    \begin{equation}
       \tilde\mu(\Delta \mu) = m_\text{F277W}^\text{obs} - m_\text{F277W}^\text{temp} + \Delta \mu \, ,
    \end{equation}
    where $\Delta \mu$ is a free parameter defined as a small offset from the difference in F277W magnitudes.
    Due to the expected non-Gaussianity in the posterior distribution of $R_V$, a simple linear regression is not suited for this study. We use Markov Chain Monte Carlo (MCMC) sampling with the \texttt{emcee} package \citep{Foreman_Mackey_2019_emcee} to estimate the posterior distribution of the parameters.
    
    The likelihood function is defined as the Gaussian likelihood of the proposed dust law with respect to the residual SED ($m^\text{obs}-m^\text{temp}$). For the $i$-th galaxy, we evaluate the likelihood as
    \begin{align}\label{eq:likelihood}
        \centering
        \nonumber &\log\mathcal{L}(\mathrm{data}_i, \mathrm{template}_i \ |\ \theta_i) \hspace{3cm}\\ \nonumber
        & \hspace{1.5cm} = (\boldsymbol{y} - \boldsymbol{F}(\boldsymbol{\theta}))^\top \mathbf{C}^{-1} (\boldsymbol{y} - \boldsymbol{F}(\boldsymbol{\theta})) \\
        & \hspace{1.5cm} \quad -\frac{1}{2} \left[ \ \ln\det(\mathbf{C}) + N \ln(2\pi) \ \right]\, ,
    \end{align}
    where $\theta$ is the set of parameters $(A_V,R_V,\Delta \mu)$, $\boldsymbol{y}$ is a vector of observed extinction at each filter (whose covariance is described by $\boldsymbol{C}$) given by
    \begin{align}
        \boldsymbol{y}_\text{filt} &= m_\text{filt}^\text{obs} - m_{\text{filt} }^\text{temp} \\
        \boldsymbol{C} &= C_\text{obs} + C_\text{temp}\, ,
    \end{align}
    and $\boldsymbol{F}(\theta) = [\boldsymbol{F}_\text{F555W}(\theta), \boldsymbol{F}_\text{F814W}(\theta), \cdots]$ is a vector of the relative extinction law to be optimized (including the adjusted distance modulus) for each filter,
    \begin{align}\nonumber
         \boldsymbol{F}_\text{filt}(\theta) &= A_\text{filt}(A_V,R_V) - A_\text{F277W}(A_V,R_V)\\
         &\hspace{4.3cm} + \tilde\mu(\Delta \mu)\ .
    \end{align}
    The equation above cancels the extinction at F277W: $F_\text{F277W}(\theta) = \tilde\mu(\Delta \mu)$, so that the inferred $\Delta \mu$ is independent of $A_V$.

\begin{figure}
        \centering
        \includegraphics[width=\linewidth]{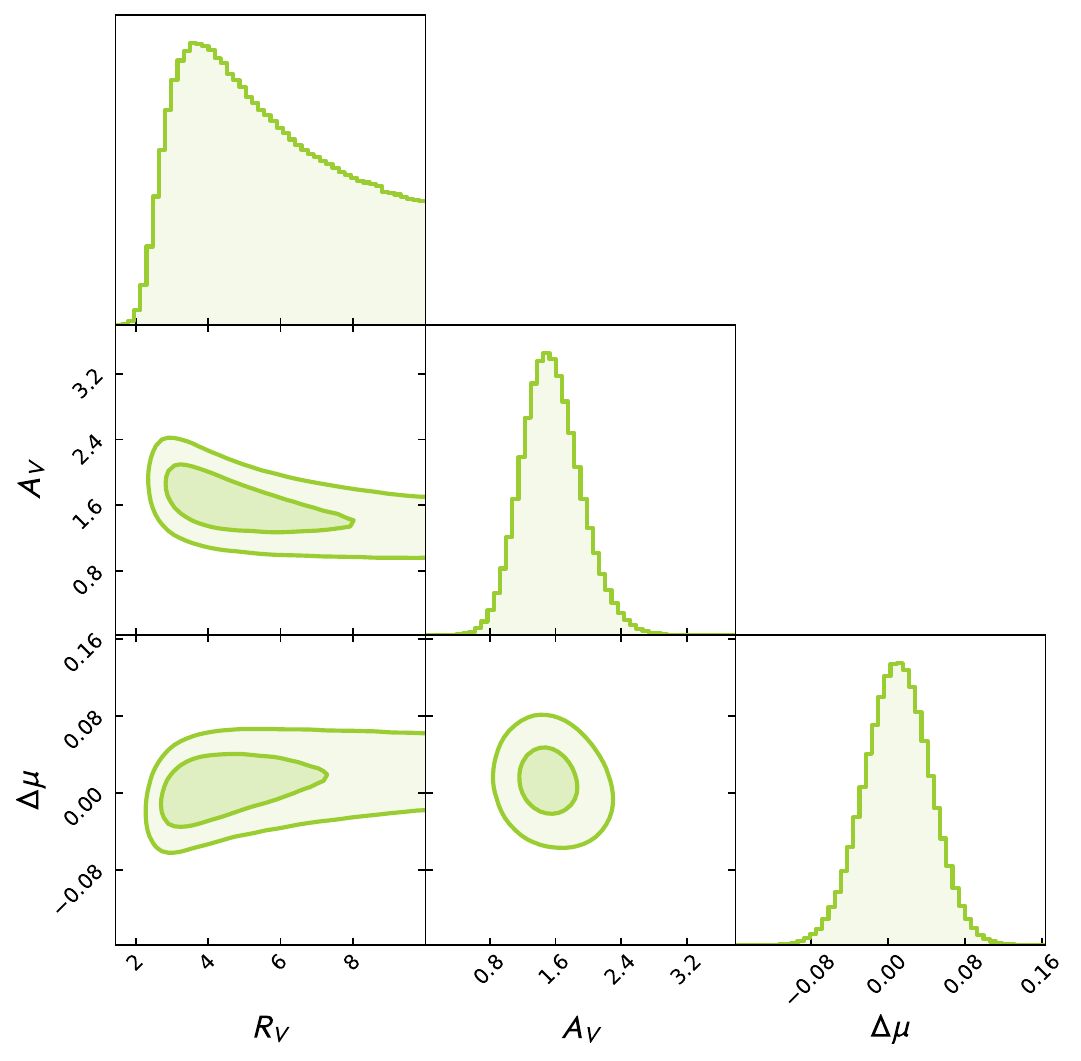}
        \caption{An example of a three-parameter fit for a single galaxy (\texttt{g004}). Small correlations between each parameter make it a better choice to fit all galaxies simultaneously, rather than combining the posterior probability density functions of $R_V$ later.}
        \label{fig:corner_single}
    \end{figure}

    For our analysis, we have 10 background galaxies, each of which yields a non-Gaussian correlation between $R_V$, $A_V$, and $\Delta \mu$ if fit individually (see Fig.~\ref{fig:corner_single}).
    Since our goal is to estimate the mean $R_V$ of NGC~5584, this makes simultaneous fitting of all galaxies a better choice. Doing so not only accounts for possible small correlations of $A_V$ values between galaxies, but also allows us to obtain an accurate picture of $A_V$ and $\Delta \mu$, as better constraints on $R_V$ improves constraints on $A_V$ and $\Delta V$.
    We therefore evaluate the joint likelihood
    \begin{equation}
        \log\mathcal{L}(\Theta) \propto \sum_i \log\mathcal{L}(\mathrm{data}_i, \mathrm{template}_i \ |\ \theta_i)\, , 
    \end{equation}
    where $\Theta$ is the combination of parameters with a single extinction law to be parsed into individual parameters for each galaxy: 
    \begin{equation}
        \Theta = [R_V,\ A_V^0, \cdots, A_V^N, \ \Delta \mu^0,  \cdots, \Delta \mu^N]\ .
    \end{equation}

    The prior is chosen to be uniform within the physically plausible range of the parameters (i.e., top-hat prior) in the following ranges:
    \begin{align*}
        A_V &\sim \mathcal{U}(0,4)\\
        R_V &\sim \mathcal{U}(0.5,10)\\
        \Delta \mu &\sim \mathcal{U}(-0.5,0.5)\, .
    \end{align*}
    The range of $R_V$ is chosen to be wider than what we physically expect. With the Rayleigh scattering limit of $R_V \approx 1.2$ \citep{Draine_2003_dust_review}, $R_V\approx 0.5$ is almost certainly not physical, and this allows us to check for systematic issues in the fitting procedure if the result points to or beyond such limit.
    
    We validate the fitting process and characterize necessary corrections (bias and systematic uncertainty) through tests described in Appendix~\ref{appendix:analysis_tests}.

\section{Results} \label{sec:results}

    \begin{deluxetable}{lrrrrr}
        \vspace{3mm}
        \tablehead{\colhead{ID} & \colhead{$A_V$} & \colhead{$\sigma_{A_V}$} & \colhead{$\Delta \mu$}& \colhead{$\sigma_{\Delta \mu}$} & \colhead{$\log\mathcal{L}$}\\
        & (mag) & (mag) & (mag) & (mag) & }
        \startdata 
        g004 & 1.68 & 0.35 &  0.003 & 0.033 & 7.92 \\
        g095 & 1.20 & 0.60 &  0.053 & 0.036 & -0.32 \\
        g144 & 1.03 & 0.46 &  0.035 & 0.038 & 5.34 \\
        g198 & 1.30 & 0.79 & -0.022 & 0.047 & 3.42 \\
        g226 & 0.89 & 0.55 &  0.033 & 0.035 & 2.65 \\
        g232 & 1.46 & 0.50 &  0.017 & 0.041 & 6.04 \\
        g260 & 0.43 & 0.31 &  0.037 & 0.032 & 5.86 \\
        g281 & 1.23 & 0.55 & -0.017 & 0.049 & 0.07 \\
        g310 & 0.33 & 0.25 &  0.030 & 0.032 & 8.08 \\
        g321 & 1.43 & 0.67 &  0.017 & 0.041 & 6.68 \\
        \enddata
        \caption{The list of our target galaxies, their aperture magnitudes with Petrosian index 0.5, and the result of extinction-law fits. }
        \label{tab:fit_results}
    \end{deluxetable}

\begin{figure}
        \centering
        \includegraphics[width=\linewidth]{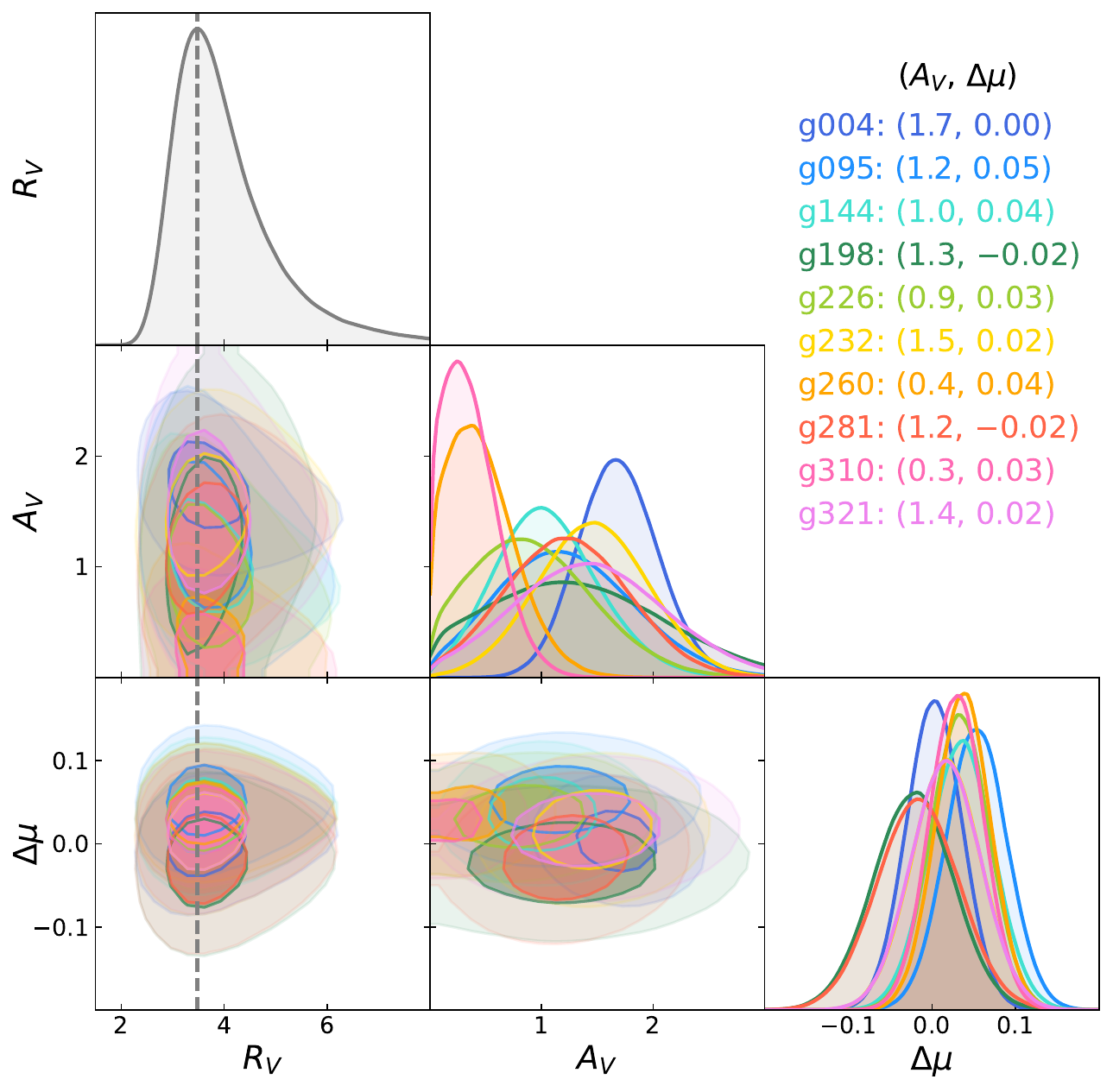}
        \caption{A simplified corner plot for the 21-parameter joint fit result for our 10 background galaxies. The foreground extinction-law parameter $R_V$ is shared across all galaxies, and the total extinction and the magnitude offset $(A_V, \Delta \mu)$ are fit to each galaxy. Note that, since this is a simultaneous fit, the posterior distribution of $R_V$ is identical for all galaxies displayed.}
        \label{fig:gtc}
    \end{figure}

    We fit all of our target galaxies using the method described in Section~\ref{sec:analysis}.
    A simplified corner plot of the posterior is shown in Figure~\ref{fig:gtc}, and the best-fit results for individual total extinction values are shown in Table~\ref{tab:fit_results}.

    The posterior samples of $R_V$  (Fig.~\ref{fig:gtc}) exhibit a strongly skewed distribution in contrast to the clean normal distribution for the total extinction $A_V$ and the photometry offset $\Delta \mu$. This non-Gaussianity --- an acute drop toward the smaller $R_V$ (left) side and a long tail toward the larger $R_V$ (right) side --- is likely due to the nonlinear relation between the value of $R_V$ and the extinction in each individual band (see Sec.~\ref{sec:discussion} for discussion).
    
    The photometry offset term $\Delta \mu$ provides an important check for the overall consistency of the photometry and/or the fitting procedure. The mean $\Delta \mu$ value should be within or approximately $1\sigma$ from zero, and a larger deviation could indicate that the estimated intrinsic SED (combined template; Eq.~\ref{eq:mag_temp}) does not match the observation. Our likelihood function (Eq.~\ref{eq:likelihood}) is formulated such that it automatically accounts for these cases by deweighting SED-mismatched galaxies upon calculating the log-likelihood based on the degree of mismatch, in addition to the size of the error bars in each data point. We confirm that all of our results are consistent with this expectation in Table~\ref{tab:target_galaxies}. The effect of SED mismatch on our analysis results is discussed in Section~\ref{sec:discussion}.

    \begin{figure*}
        \centering
        \includegraphics[width=\linewidth]{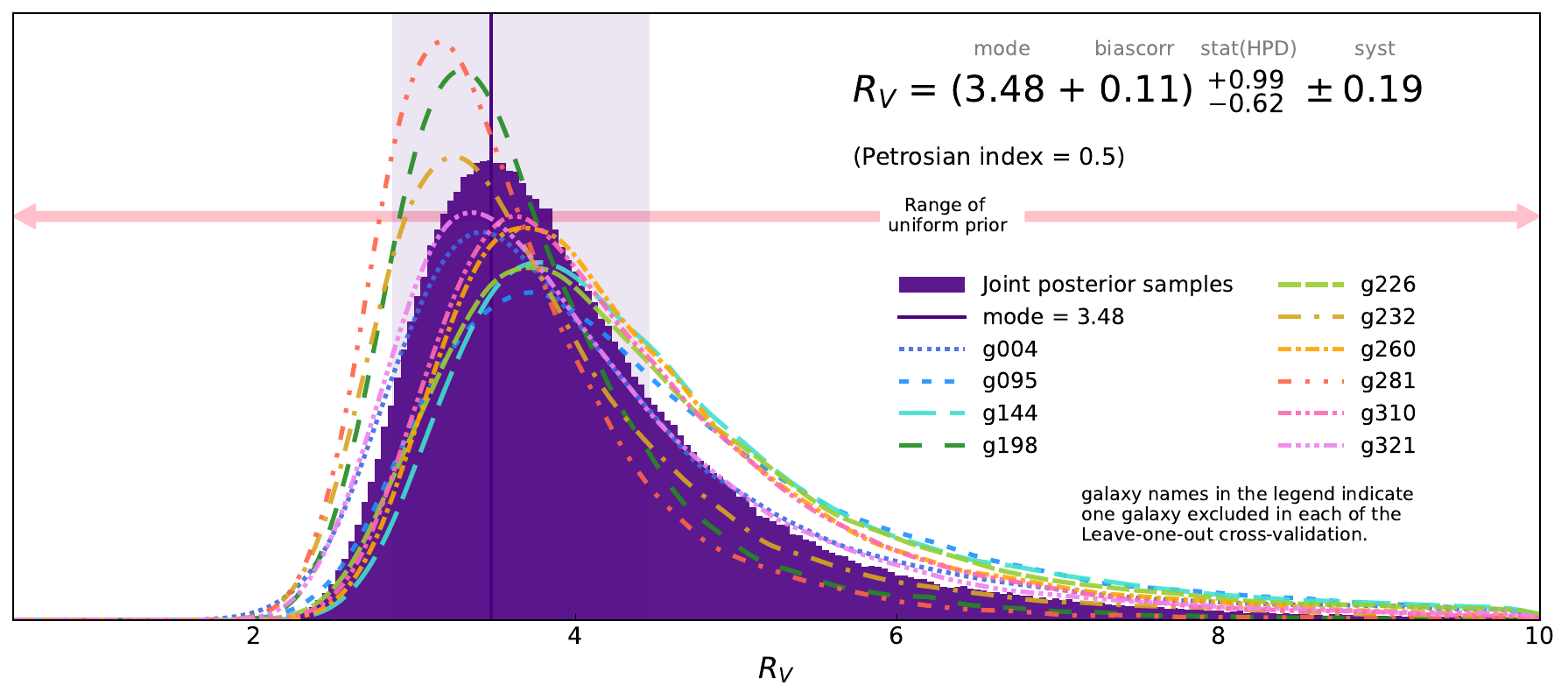}
        \caption{Our main result: the optical dust extinction law $R_V$ for NGC 5584, fitted by evaluating 10 background galaxies. Petrosian index 0.5 for the photometry is used as our baseline result, whose posterior distribution is shown as a dark histogram. Each thin curve indicates the posterior distribution of Leave-one-out cross-validation (LOOCV) runs obtained by excluding one galaxy from the dataset and repeating the analysis.}
        \label{fig:RV_result}
    \end{figure*}
    
    The result of the dust extinction-law analysis is shown in Figure~\ref{fig:RV_result}. The joint probability distribution function has the mode (peak) of $R_V = 3.48$ and median of $R_V = 3.81$, and the posterior distribution has the mean value of $R_V = 4.08$, due to the skewed profile toward the larger $R_V$ side. Following our validation test results and discussion in Appendix~\ref{appendix:analysis_tests}, we use the mode as the best-fit value, and we determine the uncertainty size by the highest probability density (HPD) interval.
    Figure~\ref{fig:RV_result} also shows the posterior distribution we obtained from the leave-one-out cross-validation (LOOCV), for which we exclude one galaxy and run the joint fit with the remaining 9 galaxies to investigate the effect of each galaxy. We see an even distribution of posterior mode below/above our baseline result without outliers, which confirms our baseline result is a well-balanced mixture of all galaxies.

    We further include additional systematic corrections, including $R_V$-dependent bias of $+0.11$ (Appendix~\ref{appendix:analysis_tests}), systematic uncertainty in $R_V$ after the bias correction ($\pm 0.01$; Fig.~\ref{fig:fitting_validation}), and a systematic uncertainty to account for the possible aperture mismatch ($\pm 0.19$; see discussion in Sec.~\ref{sec:discussion}).
    Our final result of the extinction-law fit for NGC~5584 is therefore
    \begin{equation}
        \bm{R_V^\textbf{NGC5584} = 3.59\ ^{+0.99}_{-0.62}\ (\textbf{stat}) \pm 0.19\ (\textbf{syst})} \, .
    \end{equation}
    This value is consistent with the MW-like extinction of $R_V=3.1-3.3$ and strongly rejects the steeper extinction law $R_V \approx 2$ by $>2\sigma$. We discuss the implications of this result and the future outlook in Section~\ref{sec:conclusion}.

\section{Discussion} \label{sec:discussion}

    \bfit{Non-Gaussianity of $\bm{R_V}$.} In this study, we used the conventional parameterization of the dust law with $A_V$ and $R_V$. This was a reasonable choice since it allows direct and intuitive comparison with the literature, and our shortest wavelength F555W is near the $V$ band, making it easier to interpret the results. However, the relation between $R_V$ and the extinction in magnitudes (for fixed $A_V$) is highly nonlinear. For instance, the extinction near the F814W band changes by $\sim0.8$ mag between $R_V$ of 2.0 and 3.0. The subsequent effect as the value of $R_V$ gets larger becomes increasingly small, making the difference between $R_V=5$ and $R_V=6$ be $< 0.1$ mag, for example. Although this nonlinearity exists \textit{by definition} of $R_V$, it poses challenges in statistical analysis of the results, owing to the non-Gaussianity of the posterior distribution. In future studies, we intend to try other parameterizations and discuss the effect this has on the analysis.

    \bfit{Aperture matching between photometry and spectroscopy.}
    Unlike \cite{Brown_2014_SED_atlas}, performing matched-aperture photometry is not realistic with our data: ground-based spectra have a seeing-dominated spatial profile, while space telescope images have a significantly sharper PSF. 
    There are two ways the aperture size can affect the measured SED in our work: 
    (i) galaxies with a large color gradient near the aperture could have different SEDs within the enclosed area (a physical effect), and 
    (ii) larger apertures failing to calculate the background level. Since cutout size of each galaxy image is fixed, larger aperture may lead to less statistics on the background level. Since the background level is measured on the galaxy-subtracted image, this effect may be further enhanced when galaxy subtraction is imperfect.
    Both of these effects can cause the mismatch of observed SEDs from the identified template SED, even if the spectral similarity analysis successfully computes the correct intrinsic SED.

    The effect of the aperture size on our derived SEDs is shown in Figure~\ref{fig:aperture_effect}. Overall, the deviation is within the size of the error bars, and the two galaxies showing significant deviations (\texttt{g095} and \texttt{g281}) are the two most deweighted galaxies in the calculation, indicating that the joint likelihood formula (Eq.~\ref{eq:likelihood}) successfully identified the mismatched SEDs and deweighted them in the calculation. Visual inspection of the cutout (Fig.~\ref{fig:RGB_cutouts}) suggests that the large deviation in \texttt{g095} is likely due to reason (ii) above, while the smaller (but more significant toward smaller aperture) deviation is due to reason (i) above for \texttt{g281}.
    
    In this work, we consider the systematic effect of the aperture size to be very small. We only use the absorption features in the spectrum when we evaluate the similarity to templates (Sec.~\ref{eq:template-weight}) by masking known emission lines at the detected redshift. Such features are more present in older populations, which are more dominant near the center of each galaxy. 
    
\begin{figure}
        \centering
        \includegraphics[width=\linewidth]{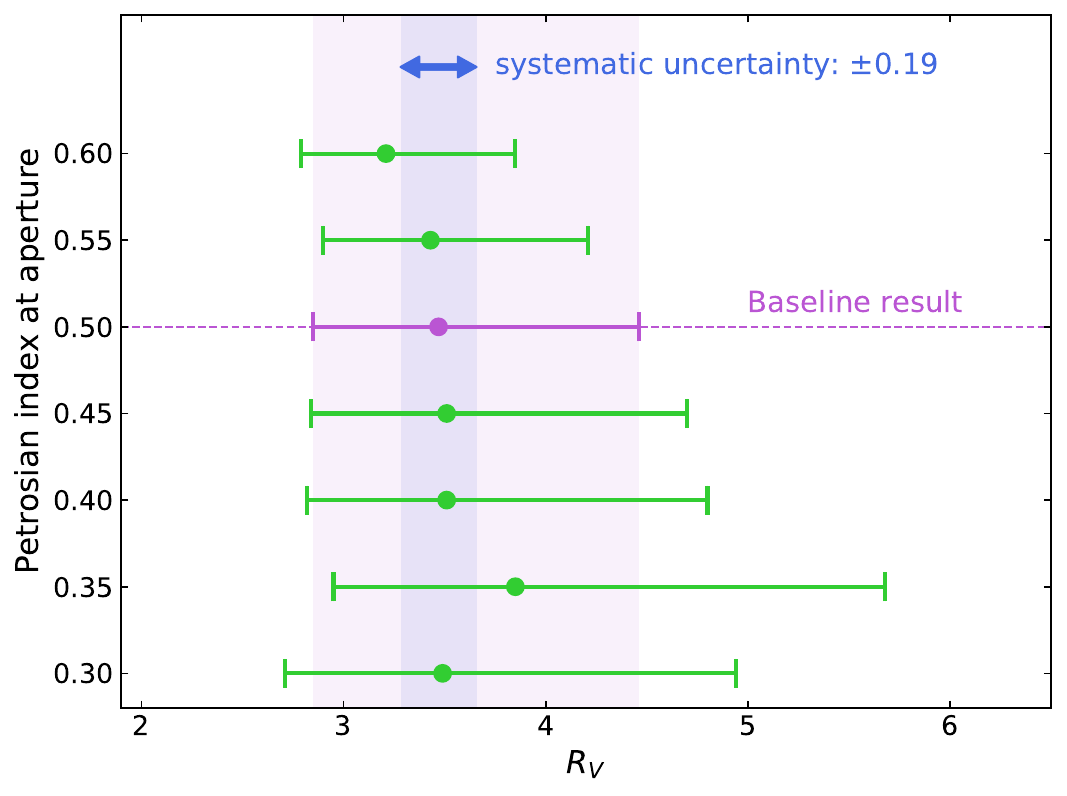}
        \caption{Evaluating the systematic uncertainty due to the aperture size by repeating the end-to-end analysis at different Petrosian indices. While we discuss that the effect is negligible in Sec.~\ref{sec:discussion}, we take the standard deviation of $R_V$ mode values (shown as the inner colored range) as the systematic uncertainty for a conservative measure.}
        \label{fig:petro_RV_endtoend}
    \end{figure}

    Still, we make a conservative measurement of the systematic uncertainty due to the aperture size selection in our final analysis. We repeat the end-to-end analysis at various aperture sizes at different Petrosian indices. For each run, we evaluate the photometric uncertainty and covariance (Fig.~\ref{fig:phot_bias_all}, Fig.~\ref{fig:corr_matrix_data}), run the validation test to identify the best configuration (Appendix~\ref{appendix:analysis_tests}), and evaluate our joint likelihood to obtain the posterior distribution of $R_V$. The mode and HPD intervals are shown in Figure~\ref{fig:petro_RV_endtoend}. Four out of six additional measurements are within $\pm0.05$ of the baseline result, and all deviations are within $0.3\sigma$ of the baseline result, showing good consistency. We take the standard deviation of all measurements and define it to be our systematic error ($\pm 0.19$). This is a conservative estimate of the systematic error and a fraction of this value may include statistical error; we aim to further investigate and reduce the size of systematic error in future projects.

    \bfit{Selection of background galaxies.} For our results to be reliable, it is important that the selection is not biased --- that is, our target galaxies are observable regardless of the $R_V$ values. To test our samples against this condition, we identify the regions in $A_V$--$R_V$ parameter space for each galaxy that would make each galaxy unobservable (either by photometry or spectroscopy) through a bootstrap test and compare them against the measured values of $R_V$ and $A_V$'s (see Appendix~\ref{appendix:bias?}). We conclude that all of our samples would be observable regardless of the $R_V$ value and our results are unlikely to be biased by the selection of background galaxies.

\section{Conclusion} \label{sec:conclusion}

    We measured the dust extinction law in NGC~5584 using photometry and spectroscopy of the background galaxies. For photometry, we developed and tested the tool \sphot, which removes the major stellar sources from the images and provides clean photometry of the target galaxies. We processed \textit{HST}/WFC3 and \textit{JWST}/NIRCam images of selected background galaxies with \sphot, and measured the SED using aperture photometry, at a fixed Petrosian index.
    Extensive testing using artificial galaxy images with the foreground galaxy NGC~5584  is used to validate the photometry, evaluate the crowding effect, and estimate the size of uncertainty and covariance within our \sphot-yielded photometry data.
    
    For spectroscopy, we used a custom slitmask observation with Keck/DEIMOS to obtain optical spectra of the target galaxies. The forced extraction method, which uses \sphot-cleaned images to estimate the spatial profile, were implemented and showed successful extraction of spectroscopic features. For each spectrum, we mask wavelengths with known emission lines at the redshift of the foreground galaxy NGC~5584, so the analysis following is not affected by the foreground stellar or ISM features.
    
    Using the spectroscopically determined redshift and the observed spectral features, we estimated the intrinsic SED of each galaxy. For this process we evaluated the similarity of flattened (continuum-removed, emission-line-masked) spectra between our observed spectra and \cite{Brown_2014_SED_atlas} templates. This provides an empirical, best-matching galaxy template whose internal physics is more similar to the observed spectra than other deweighted templates. 
    We then compute the linear combination of templates based on the spectral similarity, with a weight-scaling parameter $T$ that determines the ratio of the weights between best-matching and least-matching templates. Our tests showed that such methods can successfully recover the extinction-law parameter $R_V$ under an appropriate $T$ value.
    
    We then simultaneously fit the extinction curves to all 10 background galaxies. The analysis determines the extinction law of NGC 5584 to be
    $R_V = 3.59^{+0.99}_{-0.62} \ (\text{stat}) \pm\ 0.19 \ (\text{syst})$.
    This value is consistent with a MW-like extinction law, as predicted by its galaxy type (large spiral, star-forming galaxy similar to the MW). The posterior distribution strongly disfavors a steeper extinction curve (e.g., $R_V \approx 2$), despite multiple studies of the light curve of SN~2007af, an SN~Ia in NGC~5584, pointing toward it ($R_V=2.11^{+0.55}_{-0.48}$, \citealt{phillips_2013_extinction}; $R_V=1.8^{+0.7}_{-0.4}$,\citealt{Burns_2014_SNColor}).
    While the HPD-based uncertainty suggests $\sim 2.5\sigma$ deviation from $R_V=2$, the non-Gaussian posterior shows a further drop-down of probability toward the smaller $R_V$ value. With added noise of $\pm0.19$, the posterior distribution has $\sim0.02\%$ of the samples below $R_V\le 2$. With the shift of $+0.11$ due to the bias correction in $R_V$, that fraction reduces to $0.005\%$. This is equivalent to a 3.5--4.0$\sigma$ deviation from $R_V \approx 2$, which SN~Ia statistics predict. The same trend can be seen in all of the analysis variants (Fig.~\ref{fig:aperture_effect}).

    \begin{figure}
        \centering
        \includegraphics[width=\linewidth]{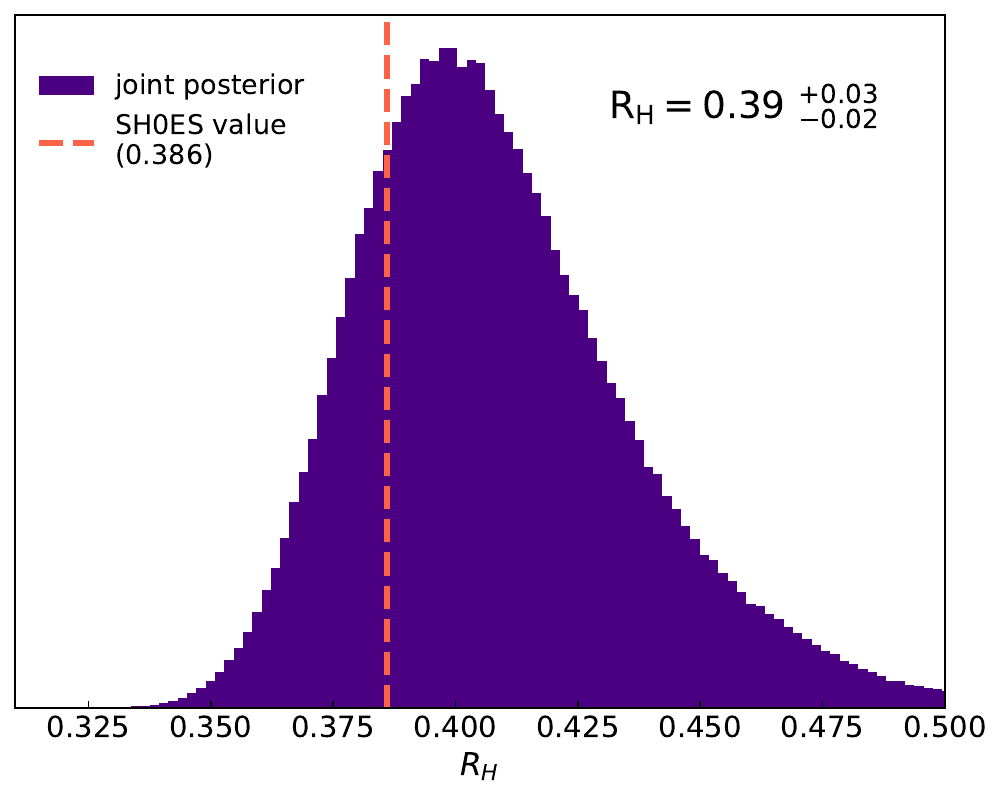}
        \caption{Our result in $R_H$, an extinction-law parameter in the NIR (see \citealt{Riess_2022_SH0ESmain} for definition).}
        \label{fig:RH_result}
    \end{figure}
    
    At NIR wavelengths, using the \cite{Fitzpatrick_1999} dust law and \cite{Riess_2022_SH0ESmain} definitions, this result corresponds to the NIR extinction slope of $R_H = 0.39^{+0.03}_{-0.02}$ (Fig.~\ref{fig:RH_result}). \cite{Mortsell_2022_H0revisited, Mortsell_2022_H0dust} have previously suggested that the possible mismatch of extinction laws between geometric anchors and Cepheid--SNIa hosts can change the Hubble constant by $\sim 1$~km~s~Mpc$^{-1}$ if the NIR extinction law $R_E$ (which is similar to the $R_H$ value) is lower than the MW-like value of $R_H = 0.386$ by $\sim 0.1$. Another argument made by the authors was that a significant variation of the extinction law, if larger than discussed by \cite{Riess_2022_SH0ESmain}, could further enlarge the size of the uncertainty from the original measurements. Our result exhibits no evidence for those discussed possibilities, as the deviation from the SH0ES value is within $0.4\sigma$. 
    A further point to note is that the individual extinction-law fits based on Cepheids presented by \cite{Mortsell_2022_H0dust} place NGC 5584 nearly at the lowest value of $R_E < 0.3$, and our results are $> 4\sigma$ away from such an extinction curve.

    While our present conclusion is for only a single galaxy and thus has no statistical constraint on the \textit{mean} extinction law of all SN~Ia (or at least SH0ES) hosts, a result consistent with a MW-like extinction law is indeed an interesting contrast to the SN~Ia-based measurements of $R_V \approx 2$. If other SH0ES galaxies are found to show comparable results, this could imply that the extinction for SNe~Ia comes from dust other than the ISM, such as circumstellar material around SNe. 
    Therefore, further measurements of dust extinction laws similar to this study, but on different and a larger set of SH0ES galaxies, are eagerly anticipated.
    
    We are currently continuing observations to obtain spectra of background galaxies for other SH0ES galaxies using Keck/LRIS and Magellan/IMACS. Additionally, a follow-up program with \textit{HST}/WFC3 (GO-17743, PI A. G. Riess) is being conducted to provide photometry of SH0ES galaxies, including NGC~5584, in three additional optical bands. These data will allow us to use the technique established in this work and repeat the analysis over many other SN~Ia hosts.
    We expect upcoming projects to provide an empirical understanding of the dust extinction in SN~Ia host galaxies and possibly reveal the underlying phenomena for the observed SN~Ia extinction. Such results will allow us to assess the true extent, or lack thereof, of the systematic effects dust extinction laws have on the local cosmological measurements and contribute toward tightening the H$_0$ measurement.

\begin{acknowledgements}
We thank our group members for their extensive support: Siyang Li for discussion of photometry, Wenlong Yuan for coordinate calibration, Louise Breuval for remarks about Cepheid color laws, Javier Manniti for comments on extinction laws and spectral analysis, and Stefano Casertano for astrophysical insights provided at meetings.
We thank Steve Finkelstein and the CEERS collaboration for sharing their catalog of CEERS (JWST-ERS-1345) galaxies prior to publication.
Y.S.M. acknowledges Stefan Arsenau for valuable discussions of the data reduction with Pypeit. Y.S.M. is grateful to Yuanze Luo for sharing knowledge of the galaxy SED and providing some of the most essential advice that pushed this project forward.
Y.S.M. thanks Alex Ho for providing help on work environments and Ruoxi Wang for proofreading and plotting advise.

A.V.F.’s research group at UC Berkeley acknowledges financial assistance from the Christopher R. Redlich Fund, Gary and Cynthia Bengier, Clark and Sharon Winslow, Alan Eustace (W.Z. is a Bengier-Winslow-Eustace Specialist in Astronomy), William Draper, Timothy and Melissa Draper, Briggs and Kathleen Wood, Sanford Robertson (T.G.B. is a Draper-Wood-Robertson Specialist in Astronomy), Heidi Gerster, Tim and Judi Hachman, Rand and Ana Morimoto, Laura Sawczuk and Luke Ellis, and numerous other donors.  

The W. M. Keck Observatory is operated as a scientific partnership among the California Institute of Technology, the University of California and NASA; the observatory was made possible by the generous financial support of the W. M. Keck Foundation. We acknowledge the excellent assistance of the staff at Keck Observatory.

\end{acknowledgements}

\appendix

\section{SPHOT Image Processing} \label{appendix:photometry}
    Here we describe the image-processing procedure with \sphot.
    Before processing the photometry data and subsequent testing with the artificial dataset, we first prepare a pixel-aligned and reprojected image, in the unit of surface brightness (MJy/Sr). The format of this data product follows the \texttt{CEERS} survey (and their rereduced \texttt{CANDELS-EGS} data), which we use for the testing (see Sec.~\ref{sec:phot_testing} and Appendix~\ref{appendix:artificial_galaxy_test}). We cross-calibrate the coordinates to the base filter F150W, and reproject the images to a scale of $0.03''$/pix, facing north up.
    Cutouts in each filter are then processed with \sphot (see Table~\ref{tab:sphot_algorithm}), which generates the background-subtracted, gradient-corrected, and PSF-subtracted image of the target galaxy (\texttt{S}). Below we describe the key steps to generate intermediate images (\texttt{A}-\texttt{F}). Further details can be found in the documentation\footnote{https://sphot.readthedocs.io/en/latest/}.
        
    \paragraph{(\texttt{A}) Raw cutout} 
    This is the untouched, ``raw" cutout near the target galaxy, in the surface brightness units MJy/Sr. The initial estimation of the coordinates and size of the galaxy is performed by two weighted one-dimensional (1D) Gaussian profiles fit to the flux profile summed over the horizontal and vertical axes. 
    The size of the galaxy is estimated from the FWHM of the Gaussian fit, and the coordinates are estimated from the peak of the Gaussian fit. \sphot then cuts out the provided images with the estimated size and coordinates to create \texttt{A}. 

    \paragraph{(\texttt{S}) Science image}
    The photometry-ready image \texttt{S} is the main product of \sphot. This is initially generated from the raw image \texttt{A} by simply removing the initial guess of the background level, and subsequent iterations update this image through PSF subtraction and sky (gradient) modeling (\texttt{S$_\text{new}$ = A - D - F}), and bad pixels identified from the grand residual (``sky-only") image \texttt{E} are masked. 
    
    \paragraph{(\texttt{B}) \sersic model} 
    An analytical model of the galaxy surface brightness profile is then fit to the image \texttt{S} from the previous iteration. The model we use is \texttt{PSFConvolvedModel2D} in the package \texttt{Petrofit} \citep{Geda_2022_petrofit}. Instead of using the default, Petrosian-profile-based prior, which requires a clean background, \sphot performs a gradient-free numerical fit to minimize the residual, using the FWHM of the Gaussian profile (described above for \texttt{A}) as an initial guess.

    \paragraph{(\texttt{C}) \sersic residual (``star-only" image)} 
    The best-fit \sersic model \texttt{B} is subtracted from the unprocessed data \texttt{A} to create the \sersic-residual image $\mathcal{C}$.

    \paragraph{(\texttt{D}) Stellar PSF scene} 
    A multi-object PSF scene (with PSF generated by \texttt{WebbPSF}\footnote{https://webbpsf.readthedocs.io/en/latest/}) is fit to the \sersic residual \texttt{C}. We use the DAO star-finding algorithm implemented in the \texttt{Photutils} package \citep{Bradley_2022_photutils} to identify sources. The initial PSF fit may struggle to cleanly identify stellar PSFs near the galaxy owing to the poor \sersic fit (which was performed on the raw image \texttt{A}), but the later iterations after improving the \sersic fit (to the cleaner, PSF-subtracted image \texttt{S}) will provide a better PSF model. The PSF fit is performed by the \texttt{IterativePSFPhotometry} function in the \texttt{Photutils} package, which is a Python implementation of the DAOPHOT algorithm \citep{Stetson_1987_DAOphot}. 
    
    \paragraph{(\texttt{E, M}) Grand residual and bad-pixel mask}
    After fitting the \sersic model and PSF scene, \sphot also generates a grand-residual image as \texttt{E = A - B - D}. This image should be flat, free from any identifiable objects (e.g., galaxy or stellar PSF) under ideal conditions, and is a representation of the sky level. Any significant deviation from the mean in this image can be therefore flagged (bad-pixel mask \texttt{M}). Flagged pixels could indicate the failed PSF fit, arm-like or ring-like structures in the galaxy which the \sersic profile cannot model, or other errors that occurred locally during the analysis. Flagged pixels are masked in the PSF residual image $\mathcal{A}'$ to avoid biasing the \sersic profile fit and reduce the contamination of the photometry.

    \paragraph{(\texttt{F}) Sky model}
    In each iteration we update the ``sky model" using the grand-residual image \texttt{E}. The image \texttt{E} is in theory nearly free of any identifiable objects, and to further reduce the chance of biasing the sky model with ill-subtracted objects, we apply a ring-median filter (\texttt{Ring2DKernel}; \citealt{astropy_2022}) to the image before fitting the model.
    The sky model is restricted to be a first-order polynomial (i.e., linear gradient with a uniform offset), which is important as any higher order will become degenerate with and bias the \sersic model.

\section{Photometry: Artificial Galaxy Tests} \label{appendix:artificial_galaxy_test}
    We test the performance of \sphot and the aperture photometry by artificially injecting galaxy images on the disk of NGC~5584 and comparing the recovered photometry with the expected value.
    
    \subsection{Test Data and Method}
    The \texttt{CEERS} survey \citep{Finkelstein_2023_CEERS_keyI} provides the perfect dataset for testing, as their data products contain the pixel-aligned images of galaxies across a number of filters, ranging from UV to NIR (\texttt{HST} images taken from \texttt{CANDELS-EGS} survey; \citealt{Stefanon_2017_CANDELS_EGS}). As mentioned previously in this section, our \textit{HST} and \textit{JWST} images are prepared in the pixel scale and units to make them compatible with \texttt{CEERS} data-release images.
    We use \texttt{CEERS} data release 0.5 and 0.6 to cover all available fields \texttt{NIRCam1-10}, as well as their {\it HST} counterpart data \texttt{HDR1} \citep{Koekemoer_2011_CANDELS}.
    
    In this \texttt{CEERS} dataset, the nearest filters available to our dataset for \textit{HST} and \textit{JWST} images are (\textit{ACS/WFC} F606W, F814W, and \textit{WFC3/IR} F160W) and (\textit{NIRCam} F115W, F150W, and F277W), replacing the mismatched F555W and F090W by F606W and F115W, respectively. While this makes the PSF-convolved shape of the galaxy in such filters slightly different from the true profile in the original filters, 
    our photometry method with SPHOT followed by PSF-corrected photometry through relatively small fixed apertures is insensitive to color-dependent changes in the true profile of the galaxy. This filter substitution makes the PSF corrections similar to the real data, so we can use these simulations to assess the systematic uncertainties.

    The deep-field galaxies used in this test are chosen and preprocessed so that their photometric properties resemble those of our dataset. We use an internal catalog of galaxies that are identified and measured by the \texttt{CEERS} collaboration to perform selections. Galaxies are chosen from the redshift range $0.1 < z < 0.9$, with the estimated total aperture magnitude in F814W being $m^\mathrm{AB}_\mathrm{F814W} > 25$. The redshift values used for these cuts are the mean of multiple photometric redshift estimates, and we only used objects whose redshift values from different estimators are within $20\%$ of each other.
    
    The test procedure follows three steps. (i) Randomly select a combination of \texttt{SH0ES} foreground stellar fields and \texttt{CEERS} background galaxies. Each background galaxy is repeated with three different foregrounds. (ii) Simulate an arbitrary foreground extinction (i.e., extinction at $z \approx 0$) by reducing the flux of the galaxy. We use the \cite{Fitzpatrick_1999} dust law with $R_V=3$, with a wide range of total extinction, $0 \le A_V \le 3$ mag. (iii) Process the stacked images (background + foreground) with \sphot. (iv) Perform aperture photometry, including the PSF-aperture correction, on both the raw \texttt{CEERS} cutout and the test data to compare the resulting values.

    \subsection{Test Results: Crowding and Bias}

    \begin{figure}[ht]
        \centering
        \begin{minipage}{0.48\textwidth}
            \centering
            \includegraphics[width=\linewidth]{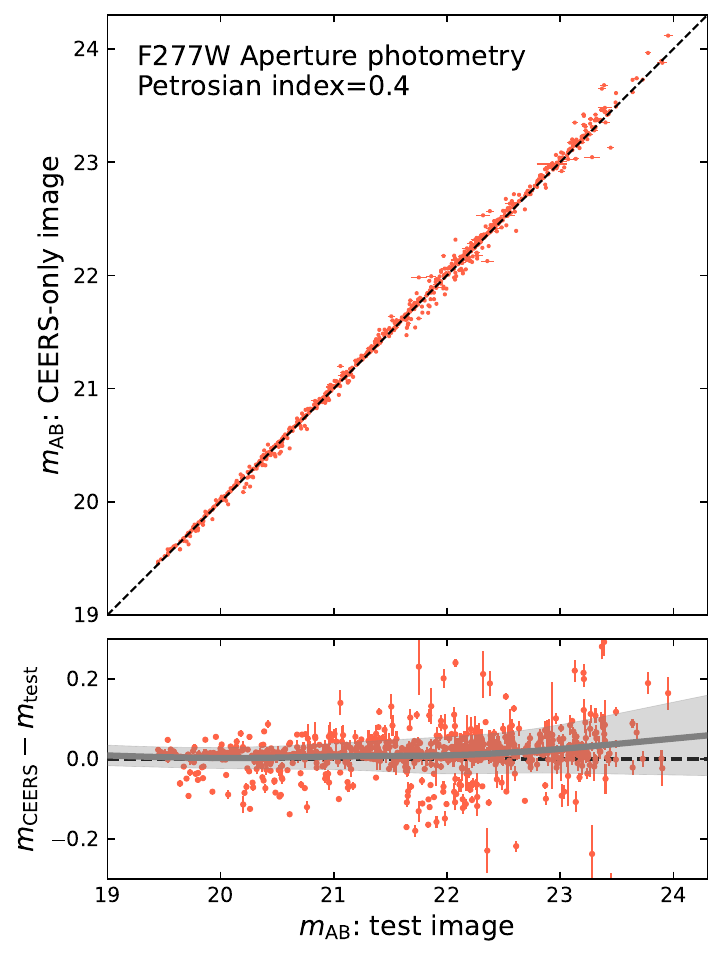}
            \caption{Artificial galaxy test result for F277W. An overall agreement at $<0.1$ mag of measured photometry between the clean \texttt{CEERS} data and \sphot-processed test data is demonstrated. A slight upward trend (gray curve in the bottom panel; Eq.~\ref{eq:phot_test_mean}) of the offset as the magnitude gets fainter is due to crowding. We use this test result to correct the measured photometry.}
            \label{fig:photometry_test_f277w}
        \end{minipage}
        \hfill
        \begin{minipage}{0.48\textwidth}
            \centering
            \includegraphics[width=\linewidth]{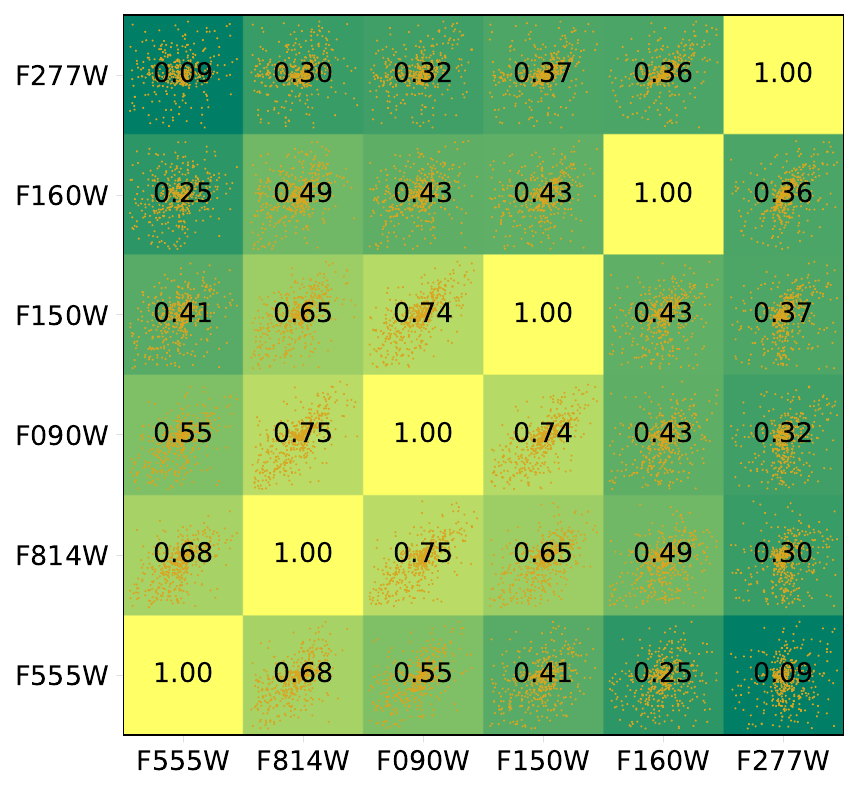}
            \caption{Correlation matrix after bias correction.}
            \label{fig:corr_matrix_data}
        \end{minipage}
    \end{figure}
    
    The test results are shown in Figures~\ref{fig:phot_bias_all} and \ref{fig:photometry_test_f277w}. Over a wide range of true magnitudes, the \sphot-processed photometry of the test data shows good agreement with deep-field photometry (e.g., upper panel of Fig.~\ref{fig:photometry_test_f277w}). 
    The offset of measured brightness from the true values helps us understand the statistical behavior of the photometry on background galaxies.
    We measure the mean trend in offset as a function of observed magnitude by fitting a spline \citep[CSAPS;][]{deBoor_1978_PGS} with iterative outlier rejection,
    \begin{equation} \label{eq:phot_test_mean}
        \mathbf{\Delta m_\text{bias}} (m_\text{obs}) = \texttt{Fit}\left(m_\text{true}-m_\text{obs}\right)\, .
    \end{equation}
    Similarly, we measure the standard deviation of the offset as a function of observed magnitude,
    \begin{align} \label{eq:phot_test_std}
        \mathbf{\boldsymbol{\sigma}_{m,\text{bias}}} (m_\text{obs}) =\ & \texttt{Fit}\left(\sqrt{\pi/2}\,
        \left|m_\text{true}-m_\text{obs} - \mathbf{\Delta m_\text{bias}}(m_\text{obs})\right|\right)\, ,
    \end{align}
    where $\sqrt{\pi/2}$ is the ratio between the standard deviation and the mean of absolute deviation, assuming a normal distribution of offsets at each observed magnitude.

    The measured offset ($\Delta m_\text{bias}$; hereafter photometry bias), seen in the bottom panel of Figure~\ref{fig:photometry_test_f277w} and in Figure~\ref{fig:phot_bias_all}, shows a slight trend for fainter galaxies to have their fluxes overestimated. This is the expected crowding effect \citep[discussions in, e.g.,][]{Riess_2024_JWSTcrowding} and is due to the presence of foreground stars' contamination within the aperture --- the residual stellar contamination near the centers of fainter galaxies has a proportionally larger impact on the photometry.
    The comparison of similar wavelengths with different PSFs demonstrates the effect of \sphot-cleaning: the crowding effect in F150W is considerably smaller than the F160W counterpart at similar measured magnitudes, thanks to the better PSF subtraction (i.e., \sphot can identify and remove fainter stars in F150W, which reduces the crowding effect). The same comparison can be made between F090W and F814W, where the crowding effect is more pronounced in F814W.
    We note that the mean offset is not centered at zero at brighter magnitudes in some filters, which may be the effect of the inaccurate PSF model, imperfect sky-gradient model, or statistical fluctuation (especially toward the tail of the sample distribution). For this work, we use this empirically measured photometry bias to correct the measurement (discussed more in Sec.~\ref{sec:photometry_correction}), which should eliminate the systematic effects. We aim to understand the origin of these features in future projects.
    
    The standard deviation ($\sigma_{m,\text{bias}}$; hereafter photometry scatter) increases nearly monotonically as the galaxy gets fainter, which is consistent with the expectations. Compared at a fixed observed magnitude, we observe that the scatter is larger when (a) the PSF FWHM is larger, (b) the foreground is more crowded, and/or (c) the galaxy morphology is more complex (which affects the \sersic profile fit and subtraction). 
    The scatter plateaus at $\sim0.2$ mag for F555W, possibly due to the cuts we applied at $\sigma_m \le 0.4$ mag. The same cuts are applied to the real data to ensure the consistency of this test with the science results.

    \subsection{Test Results: Covariance Matrix}

    The scatter size discussed in the previous section captures the overall uncertainty of the photometry \textit{across the whole test dataset}. This provides a good estimation of systematic uncertainties. For a \textit{single galaxy} that is measured in multiple filters, however, the deviation of measured photometry from the true value may be correlated across filters, especially if the deviation from the true value originates from the presence of the foreground stars whose locations are fixed across all filters. To fully capture the statistical characteristics of the photometry, we calculate the covariance matrix of the photometry error for each galaxy.

    Measuring the covariance directly, however, is challenging because the scatter size changes at different observed magnitudes. For each background galaxy we intend to use for science analysis, obtaining enough samples of artificial galaxy test results that have comparable brightness in each band is not realistic. Instead, we measure the \textit{correlation} matrix from the test results, which is insensitive to the absolute scale of the scatter for each filter, and combine it with the measured scatter size (Eq.~\ref{eq:phot_test_std}) to estimate the covariance matrix for the photometry of the real data.
    
    We evaluate the correlation matrix from a set of realistic, high-quality test data that pass a stricter cut than the previous analysis (small error from \sphot, brightness range similar to the NGC~5584 background galaxies, and the aperture size --- all so that the test data resemble the real data as much as possible). The deviation of the measured brightness of each galaxy from the true value is corrected for the mean bias (Eq.~\ref{eq:phot_test_mean}) and scaled by the scatter size (Eq.~\ref{eq:phot_test_std}) before evaluating the correlation so that the correlation matrix $R$ is not affected by the crowding effect and is not sensitive to data points with larger scatter. Using the measured deviation of the brightness for the $k$-th test image at the $i$-th filter, $\delta m^k_i = m_i^{k,\text{true}} - m_i^{k,\text{obs}}$, we evaluate the correlation of photometry error between the $i$-th and $j$-th filters as
    \begin{align} \label{eq:phot_covariance} 
        R_{ij} =& \frac{1}{N_\text{gal}}\sum_k^{N_\text{gal}} 
        \left[\left(\frac{\delta m^k_i - \boldsymbol{\Delta m}_{\text{bias},i}(m_i^{k,\text{obs}})}{\boldsymbol{\sigma}_{\text{bias},i}(m_i^k)}\right) 
        \cdot\left(\frac{\delta m^k_j - \boldsymbol{\Delta m}_{\text{bias},j}(m_j^{k,\text{obs}})}{\boldsymbol{\sigma}_{\text{bias},j}(m_j^k)}\right)\right]\, .
    \end{align} 
    
    The correlation matrix is shown in Figure~\ref{fig:corr_matrix_data}.
    We find a significant correlation between different filters --- when a galaxy is measured to be brighter under the presence of foreground, all filters exhibit the same trend. It is especially noteworthy that the correlation between filters at similar wavelengths is stronger than those at different wavelengths. This is due to the colors of foreground stars --- for example, stars visible (and thus affecting the photometry) in F555W are likely to be also visible in F814W, but is significantly less visible in F277W, where redder stars dominate. We use this correlation matrix to estimate the covariance matrix for each galaxy in the following section.
    
    \subsection{Applying the Correction} \label{sec:photometry_correction}
    Combined with the aperture correction (Sec.~\ref{sec:aperture_corr}), our crowding-corrected, aperture-adjusted photometry measurement for each filter is thus obtained as
    \begin{align}\label{eq:mag_obs}
        m_{\text{corr},i}^\text{obs} &= m_{\sphot,i} - \Delta m_{\text{aper},i} + \boldsymbol{\Delta m}_{\text{bias},i}(m_{\sphot,i} - \Delta m_{\text{aper},i})\, .
    \end{align}

    The uncertainty and the relation between each filter's data are described by the covariance matrix as
    \begin{equation} \label{eq:phot_cov}
        C_\mathrm{phot} = D_\text{bias}RD_\text{bias} + D_\text{\sphot}\, ,
    \end{equation}
    where $R$ is the correlation matrix (Eq.~\ref{eq:phot_covariance}),
    $D_\text{bias}$ is the diagonal matrix of the scatter size from the artificial galaxy test (Eq.~\ref{eq:phot_test_std}), 
    and $D_\text{\sphot}$ is the diagonal matrix of the photometry,
    \begin{align}\nonumber
        D_\text{bias} &= \begin{bmatrix}
            \boldsymbol{\sigma}_{\text{bias},1}(m_1) & 0 & \cdots & 0 \\
            0 & \boldsymbol{\sigma}_{\text{bias},2}(m_2) & \cdots & 0 \\
            \vdots & \vdots & \ddots & \vdots \\
            0 & 0 & \cdots & \boldsymbol{\sigma}_{\text{bias},6}(m_6) 
            \end{bmatrix}\, , \\
        D_\sphot &= \begin{bmatrix}
            \sigma_{\sphot,1} & 0 & \cdots & 0 \\
            0 & \sigma_{\sphot,2} & \cdots & 0 \\
            \vdots & \vdots & \ddots & \vdots \\
            0 & 0 & \cdots & \sigma_{\sphot,6}
            \end{bmatrix}\, .
    \end{align}
    We will use our photometry data ($m_{\text{corr},i}^\text{obs}$, $C_\mathrm{phot}$) in the analysis in Section~\ref{sec:analysis}.

\section{Spectroscopy: Extracting Faint Galaxies} \label{appendix:spectroscopy}

    \subsection{Initial Reduction and Sky-Foreground Subtraction}
    The data reduction is performed by a combination of \texttt{PypeIt} \citep{Prochaska_2020_Pypeit} and a custom routine. Owing to the rapid and ongoing developmental status of \texttt{PypeIt} at the time of data reduction, a few steps in data reduction were performed manually.

    We prepare the flat, wavelength-calibrated 2D CCD image (hereafter \texttt{SCI}) with the standard, automated reduction process with \texttt{PypeIt}. These individual images are stacked with a median filter.
    Focused on the target galaxies near NGC 5584 (CCD3; the second from left on the slitmask design image in Fig.~\ref{fig:observation_layout}), we estimate the mean ``sky'' model by using all pixels in \texttt{SCI} except the masked regions near objects. A major difference between the ordinary sky emission data \citep[e.g.,][]{Hanuschik_2003_skyemission} and this ``sky'' model exists in the presence of the foreground galaxy: the mean counts in each wavelength bin in our data include both the emission lines and continuum from the night-sky atmosphere but also the mean foreground spectrum (hence ``sky-foreground model''). We use this as the first iteration of the process to remove the contamination from the foreground galaxy NGC 5584.
    
    We achieve the sky-foreground modeling iteratively by (i) masking the object regions, (ii) fitting the sky spectrum with a polynomial, (iii) subtracting the sky spectrum from the object+sky spectrum, (iv) identifying objects, and (v) repeating the process until the sky spectrum converges. During this process, we apply a tenth-order polynomial model to correct wavelength calibrations in each slit. The resulting 2D image is \texttt{SUB}.

    \subsection{Forced Extraction}
    After the mean, global sky-foreground model is subtracted from the \texttt{SCI} 2D image, we reduce 2D images of individual slits to 1D spectra.
    The standard procedure for this step is optimal extraction \citep{Horne_1986_optimal_extraction}. However, our DEIMOS data are at optical wavelengths where the effect of extinction is significant, and the S/N of our data is low.  Many slits contain visible nonuniform foreground features (e.g., bright stars), and the standard optimal extraction algorithm in \texttt{PypeIt} failed\footnote{At the time of our data extraction, object detection with \texttt{PypeIt} was only possible on individual images before stacking, which made it nearly impossible to identify objects that have low S/N.} to detect many of our target galaxies. 

    Fortunately, we have \sphot-cleaned images of the target galaxy, which has a significantly reduced effect on the bright foreground objects. We take the \sphot-cleaned F277W image ($\mathcal{S}_\text{F277W}$), convolve it with the Gaussian kernel to simulate seeing, and project it onto the DEIMOS pixel scale and instrumental angle ($\mathcal{S}'_\text{F277W}$) to construct a 1D spatial profile,
    \begin{equation}
        P^\text{gal}_x \propto \sum_y M\mathcal{S}_\text{F277W}' \ ;\ \sum_x P^\text{gal}_x = 1\, ,  
    \end{equation}
    where $M$ is the slitmask. In this work, we assume the color gradient of the spatial profile is negligible. The constructed profile, $P$, is therefore an approximation of the locations and the spatial profile of the flux.
    To further account for the seeing from the ground, we convolve this profile with a Gaussian kernel with its standard deviation corresponding to the half of the seeing.

    This spatial profile allows us to also prepare the spatial weight for the local (slit-unique) sky-foreground model,
    \begin{equation}
        P^\text{sky}_x \propto \left[\left(P^\text{gal}_x\right)_\text{MAX}- P^\text{gal}_x\right] \ ;\ \sum_x P^\text{sky}_x = 1\, ,
    \end{equation}
    which is used to prepare an updated sky-subtracted slit 2D image as described below. 
    
    Similarly to the iteratively constructed spatial profile of \cite{Horne_1986_optimal_extraction}, we use these predetermined profiles to extract  spectra of the local foreground population and the sky. 
    The local sky-foreground model is
    \begin{equation}
        f^\text{sky}_\lambda = \frac{\sum_x P^\text{sky}_x \cdot \mathtt{SUB} / \sigma_\mathtt{SUB}^2}{\sum_x \left(P^\text{sky}_x\right)^2/\sigma_\mathtt{SUB}^2} \, ,
    \end{equation}    
    which is used to generate the updated sky-subtracted image $\mathtt{SUB}'$.
    The flux of the target galaxy is therefore
    \begin{equation}
        f^\text{gal}_\lambda = \frac{\sum_x P^\text{gal}_x \cdot \mathtt{SUB}' / \sigma_\mathtt{SUB'}^2}{\sum_x \left(P^\text{gal}_x\right)^2/\sigma_\mathtt{SUB'}^2}\, .
    \end{equation}
    During this process, the profiles were allowed to move along the spatial direction by a constant offset, which was optimized, to account for the pointing and coordinate calibration offset.

\section{Estimating the Intrinsic SED Based on Spectroscopic Similarity} \label{appendix:spectrum-matching}
    \textbf{\textit{Spectroscopic Similarity}} --- 
    We determine the similarity between the observed and each template spectrum by computing the $\chi^2$ value over a grid of wavelengths. Both observed and template spectra are continuum-removed, and we mask emission lines (since we are interested in the stellar population that makes up the SED rather than ISM emission). The template is redshifted to the observed galaxy's frame and scaled to match the observed flux. The $\chi^2$ value\footnote{See footnote~\ref{footnote:chi2_dont_be_confused}.} of the $i$-th template against the observed spectrum is calculated as
    \begin{equation}
    \chi^2_{i,\mathrm{DOF}} = \frac{1}{N_\text{wav}} \sum_\lambda^{N \text{wav}} \left(\frac{f_\lambda^\text{obs} - f_{i,\lambda}^\text{temp}}{\sigma^\text{obs}_\lambda}\right)^2\, .
    \end{equation}
    The smaller $\chi^2$ value corresponds to a better-matching spectrum.
    We then use this value to determine the weight for the linear combination of templates. The weight is calculated similarly to the Gaussian likelihood,
    \begin{equation} \label{eq:template-weight}
        w_i = \exp\left(-\frac{\chi^2_{i,\mathrm{DOF}}}{\langle\chi^2_{\mathrm{DOF}}\rangle T}\right)\, .
    \end{equation}

    In Equation~\ref{eq:template-weight}, the term $\chi^2_{i,\mathrm{DOF}} / \langle\chi^2_{\mathrm{DOF}}\rangle$ behaves like a standardized $\chi^2$ that represents the goodness of fit of each template to the data \textit{with respect to the average}. 
    We use this standardized $\chi^2$ instead of the absolute measurement to account for the potentially underestimated or overestimated uncertainties in the observed spectra, which can significantly affect the separation of the good and bad templates in the linear combination. By defining a constant value of the ``mean similarity," we not only rank the templates by their similarity to observation but also evaluate the overall goodness of the best-matching template compared with the average. 

    \bfit{Scaling Parameter $\bm{T}$} --- 
    The parameter $T$ normalizes the exponential function and controls the relative weight of the templates. The choice of $T$ is critical for the linear combination, as it determines the relative importance of the goodness of fit to the average. A small value of $T$ will result in the combined template that is dominated by the template that fits the data  best (i.e., most selective), while a large value of $T$ will result in the combined template that is dominated by the average of the templates (i.e., least selective), which provides no additional information.
    In our analysis, smaller $T$ (i.e., more selective) results in a smaller uncertainty, but could possibly be biased due to the smaller effect of averaging (i.e., $\chi^2 \gg 1$), and larger $T$ corresponds to a more averaged template with overestimated uncertainty (i.e., $\chi^2 \ll 1$). The choice of $T$, therefore, must be tested with simulations so that the resulting uncertainty in the combined template best describes the statistical deviation of the data from the ``true'' SED.
    We determine the best value of $T$ based on the validation test results (see next section).

    \bfit{Combined Flux} ---
    Once the weight for each template is determined, we calculate the linear combination of templates (hereafter ``combined template''). 
    The combination is performed in the observed frame  with the redshift of the target galaxy and is in the flux space,
    \begin{equation}\label{eq:flux_temp}
        f^\text{comb}_\text{filt} = \frac{\sum_l w_l \left(f_{l,\text{filt}}/f_{l,\text{F277W}}\right)}{\sum_l w_l}\, , 
    \end{equation}
    where $f_{l,\text{filt}}$ is the flux of the $l$-th template at each bandpass filter, computed in units of Jansky.
    The templates are normalized to the flux at $\sim2.7\,\mu$m since the  extinction analysis is based on the color measurement with respect to F277W.
    The combined fluxes are then converted to the AB magnitude system for the comparison with the observed photometry,
    \begin{equation}\label{eq:mag_temp}
        m^\text{comb}_\text{filt} = -2.5\log_{10}\left(f^\text{comb}_\text{filt}\right) - m^\text{comb}_\text{F277W}\, .
    \end{equation}

    \begin{figure}
        \centering
        \includegraphics[width=0.5\linewidth]{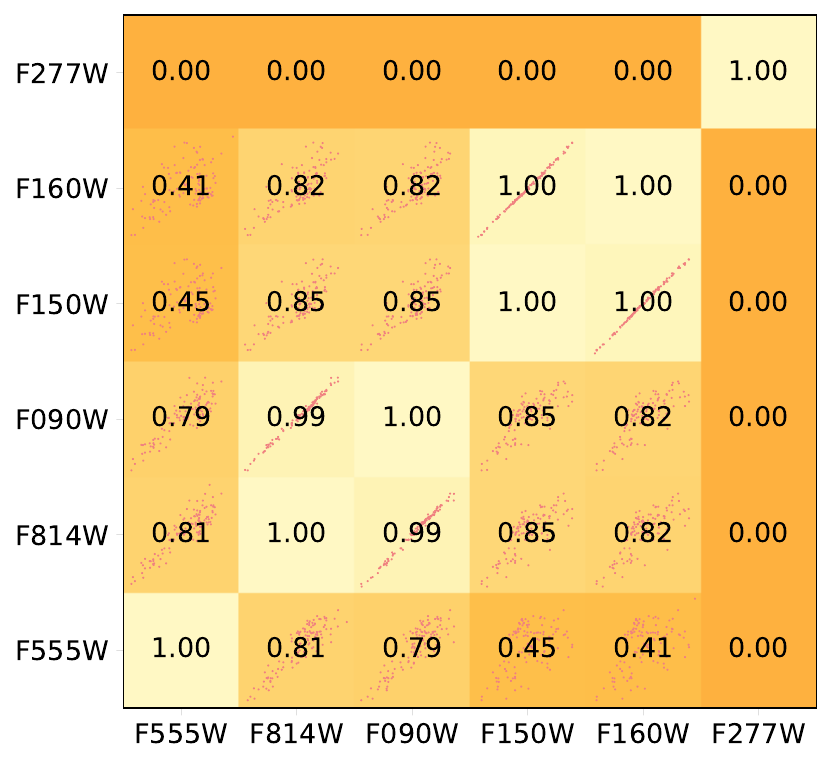}
        \caption{An example of the correlation matrix for the combined template SED. Unlike Fig.~\ref{fig:corr_matrix_data}, we evaluate the \textit{covariance} matrix directly using the weights $w_i$. This matrix is provided for an intuitive comparison with Fig.~\ref{fig:corr_matrix_data}.}
        \label{fig:corr_matrix_template}
    \end{figure}

    \bfit{Covariance Matrix} ---
    Since the synthetic bandpass fluxes in six bandpass filters $f_{l,\text{filt}}$ are sampled from a single ($l$-th), \textit{continuous} template (c.f., photometric data measured from six independent images of the target object), some neighboring filters (e.g., F150W and F160W) are bound to have nearly identical flux values. Because of this, we expect a significant covariance between filters --- the template is dimmer across all filters when the template is redder, and the template is brighter across all filters when the template is bluer. The measurement of the combined template photometry $m^\text{comb}_\text{filt}$ should be therefore associated with this covariance. We evaluate the covariance matrix of the combined template fluxes using the weights $w_i$. For the $i$-th and $j$-th filters, the covariance is evaluated as
    \begin{equation} \label{eq:cov_template_flux}
        C^\text{flux}_{\text{temp},ij} = \frac{\sum_l w_l \left(f_{l,i} - f^\text{comb}_i\right)\left(f_{l,j} - f^\text{comb}_j\right)}{\sum_l w_l}\, .
    \end{equation}
    Propagating this covariance measured in the flux scale into the AB magnitude system, we obtain the covariance matrix of the combined template photometry,
    \begin{equation} \label{eq:cov_template_mag}
        C^\text{mag}_{\text{comb},ij} = \frac{2.5^2}{m_i m_j\left(\log 10\right)^2} \ C^\text{flux}_{\text{temp},ij} + \text{diag}(0.03)\, .
    \end{equation}
    We add a $\sim 3\%$ uncertainty to the diagonal elements of the covariance matrix to account for the possible uncorrelated random noise in individual templates as a conservative measure. An example of the correlation matrix, converted to the \textit{covariance} matrix for a direct comparison against Figure~\ref{fig:corr_matrix_data}, is shown in Figure~\ref{fig:corr_matrix_template}. As expected, there is significant correlation between neighboring filters, and the covariance matrix is significantly larger than the counterpart from Section~\ref{sec:photometry_correction}.
    For simplicity and the simple comparison against the observed data ($m_\text{obs}$, $C_\text{obs}$), we refer to the combined template photometry and its covariance matrix as $(m_\text{temp}, C_\text{temp})$ in the following section.

\section{Spectral Matching: Testing and Validation} \label{appendix:analysis_tests}
    \begin{figure*}
        \centering
        \includegraphics[width=\linewidth]{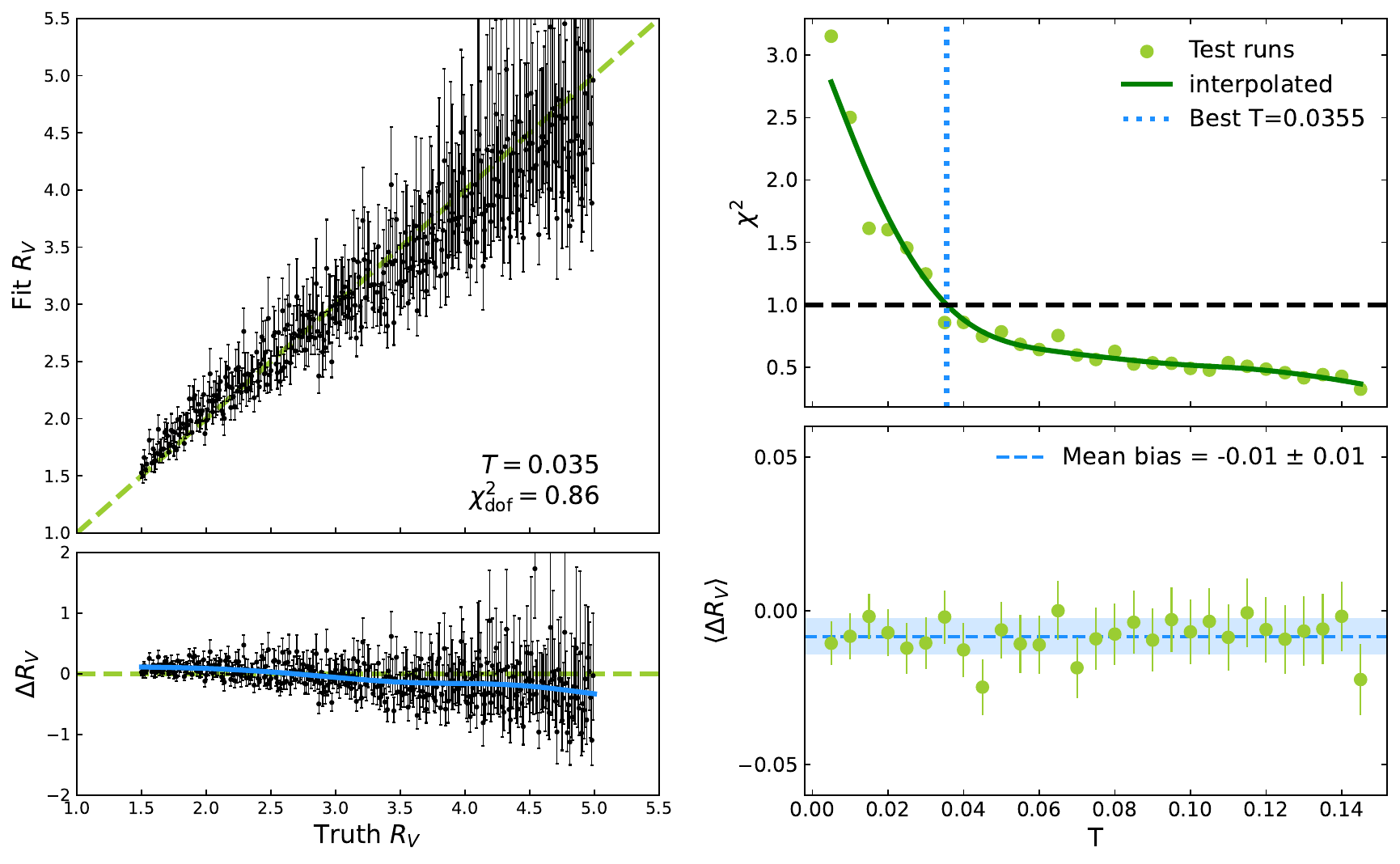}
        \caption{
        \textit{Left:} An example of a single test run at a fixed $T$ value. Each point and the associated error bar represent the 16-50-84th percentiles of the joint PDF, which are created by repeating the attenuation fit procedure to 10 randomly selected templates redshifted to $z=0.6$.
        \textit{Top right:} The overall test result. The $\chi^2$ measuring the deviation between measured and true $R_V$ is shown as a function of the relative weight parameter $T$. Each data point represents a set of tests over the range of $R_V$, as shown in Figure~\ref{fig:fitting_validation}. The best value of $T$ is determined by the interpolated profile where $\chi^2=1$.
        \textit{Bottom right}: The mean deviation of the fitted $R_V$ value to the truth value.
        }
        \label{fig:fitting_validation}
    \end{figure*}

    Our estimate of $R_V$ and the uncertainty is calculated through the procedure of (a) evaluating the spectral similarity using flattened spectra, (b) preparing the combined templates based on the similarity, and (c) jointly fitting the dust extinction parameters. The size of the uncertainty, as well as the overall accuracy, depends on the scaling parameter $T$ (Eq.~\ref{eq:template-weight}), and therefore a validation test is critically important for determining the optimal value of $T$ to produce reliable results. The validation test also provides the possible correction needed to the fitted $R_V$ value and informs us about the best statistical metrics to evaluate the posterior distribution.

    \bfit{Test Method} --- We perform the validation test by running the whole procedure (a -- c) to the simulated dataset we generate from \cite{Brown2014_templates} templates. 
    This test is performed over a grid of ($R_V^\text{\ truth}$, $T$), where each grid point produces one joint PDF of $R_V^\text{\ fit}$ as a result of fitting to 10 artificial target galaxy datasets. 
    The simulated dataset at each run (grid point) consists of 10 ``background'' galaxies, randomly drawn from \cite{Brown2014_templates} templates, and redshifted to $z=0.6$. The photometry of these galaxies is then perturbed by the noise drawn from a covariance matrix. We prepare the covariance matrix to simulate the noise by taking the mean of all 10 background galaxies' covariance matrices in  flux space. Finally, dust extinction with a unique total extinction $A_V$ (randomly drawn between $0.1\le A_V \le 3.0$ mag) for each galaxy, with the $R_V$ value corresponding to each grid, is applied. A flattened spectrum for each galaxy is prepared by removing the continuum from the optical spectrum (5300--9500\,\AA), adding $10\%$ uncorrelated noise at each wavelength, applying the redshift, and resampling the wavelengths at $8.4$\,\AA/pix resolution to match the quality of observed spectra (Sec.~\ref{sec:spectroscopy}). This flattened spectrum and the photometry are then used to fit the dust extinction law using the automated procedure (a -- c) described above.
    The result from the joint PDF is then compared to the true value of $R_V^\text{truth}$ using the statistical metrics discussed below. 
    
    \bfit{Mode, Mean, and Uncertainty} --- As discussed later, we find that the posterior distribution of $R_V$ is skewed, and there is often a significant difference between the mode and the median of the distribution. We test two metrics to define the ``best-fit'' value for each run: the mode and the median of the distribution. Similarly, we compare two metrics to define the uncertainty: the 16--84th percentile, and 68\% highest probability density (HPD) interval. The HPD interval is defined as the smallest interval that contains 68\% of the probability density, and  \cite{Chen_Shao_1999_HPD_MCMC} argue that the HPD interval is more desirable for a nonsymmetric posterior.
    Of four possible combinations of these metrics, we find that the (mode, HPD) produces the smallest bias and $\chi^2$ closer to $1$, indicating that this combination is more reliable for our analysis\footnote{We find that the mode of the posterior distribution has better agreement with the true $R_V$ value than the median. The percentile-based uncertainty tends to underestimate the size of the shorter tail while overestimating the uncertainty in the longer tail, resulting in an inconsistent behavior of $\chi^2$  depending on whether the mode (or median) is above or below the true $R_V$ value.}.

    \bfit{$\bm{R_V}$-dependent bias} --- The left panel of Figure~\ref{fig:fitting_validation} shows the result of the test at a fixed $T$ value. Over a wide range of true $R_V$ values, the fitted $R_V$ (mode) shows good agreement. A small $R_V$-dependent bias is observed toward the larger true $R_V$ values, and correcting this bias produced a much smoother $\chi^2$--$T$ profile (which we discuss below). The $R_V$-dependent bias, shown as a blue curve in Figure~\ref{fig:fitting_validation}, is evaluated by fitting a spline similarly to Equation~\ref{fig:phot_bias_all}. We then convert this trend to a one-to-one function between the fitted and true $R_V$ values, which we use to correct the fitted $R_V$ value.

    \bfit{$\bm{\chi^2_\textbf{dof} - T}$ Profile} --- The reduced $\chi^2$ values,\footnote{\label{footnote:chi2_dont_be_confused}Not to be confused with the $\chi^2$ values in Sec.~\ref{appendix:spectrum-matching}: this $\chi^2$ value measures the deviation of best-fit $R_V$ against the expected value to evaluate the result of tests, while Sec.~\ref{appendix:spectrum-matching} uses the $\chi^2$ statistic to measure the similarity between the observed spectrum and a template.} 
    $\chi^2_\text{dof}$, are shown as a function of the relative weight parameter $T$ in the top-right panel of Figure~\ref{fig:fitting_validation}. Each data point represents a set of tests over the range of $R_V$, as shown in the left panel. 
    The $\chi^2_\text{dof}$ values are calculated after correcting for the $R_V$-dependent bias using the mode and HPD interval of the posterior distribution.
    As discussed in Section~\ref{sec:template_combination}, we find that a larger $T$ value results in a less physically informative results with overestimated uncertainty (due to more averaging of templates), and a smaller $T$ value results in a selective, overfitted result with underestimated uncertainty. The best value of $T$ that produces the most statistically consistent result is determined by the interpolated profile and evaluating the $T$ value where $\chi^2_\text{dof}=1$.

    \bfit{Systematics from the Choice of $\bm{T}$ Value} --- We highlight that, after applying the bias correction, the mean residual is nearly zero, and the $\chi^2$--$T$ profile is smooth, both of which indicate that the $R_V$-dependent bias corrects most of the systematic effects in the fitting procedure. 
    The bottom panel of Figure~\ref{fig:fitting_validation} shows the mean deviation of the fitted $R_V$ value to the true value after the bias correction over the range of $T$ values. The grand mean (averaged over the $R_V$--$T$ grid) is $\langle\Delta R_V\rangle = -0.01 \pm 0.01$, which is shown as the blue line and the shaded region in the panel. We define the systematic effect due to the choice of $T$ value as this standard deviation $\pm0.01$. This result suggests that the science uncertainty in our analysis is dominated by the statistical uncertainty in the data, rather than the systematic uncertainty due to the choice of $T$ value.

\section{Are Our Samples Biased?} \label{appendix:bias?}
    We test the selection of our sample galaxies by measuring the combination of $(A_V, R_V)$ values that would make them unobservable. 
    The selection of our targets is a result of visual identification, photometry, and spectroscopy. Thus, we define the condition unobservable as the following. (a) Any of the six filters (F555W, F814W, F090W, F150W, F160W, or F277W) reach the limiting magnitude. We make a conservative estimate of the limiting magnitude at the 84th percentile of the aperture magnitude within a distribution of successful \sphot photometry from the artificial galaxy test\footnote{We consider the resulting distribution of aperture magnitudes from the artificial galaxy test as a reasonable source for estimating the limiting magnitude, since it uses the foreground image from NGC~5584 and it only includes the test runs with successful photometry.}. This corresponds to $m^\text{aper}_\text{AB} =$ (26.5, 24.7, 24.0, 23.5, 23.4, 22.9), respectively. (b) The mean optical S/N of Keck/DEIMOS spectroscopy is below 3 after binning. With our setting this corresponds to $V_\text{AB} \approx 26.5$ mag.
    
    We perform the test with the following steps. (i) Use the best-fit parameters ($A_V$, $R_V$) to deredden the observed aperture magnitude of our galaxies. (ii) Using the dereddened SED as the mean, bootstrap the photometry by drawing samples from the covariance matrix for each galaxy. (iii) Draw an arbitrary set of extinction parameters ($A_V$, $R_V$) from a uniform distribution ($A_V \sim \mathcal{U}(0,6),\ R_V \sim \mathcal{U}(1,7)$). Apply the extinction to the bootstrapped photometry and test the conditions (a) and (b) above. Keep the ($A_V$, $R_V$) sample if the galaxy is ``unobservable.''

\begin{figure*}[h]
        \centering
        \includegraphics[width=\linewidth]{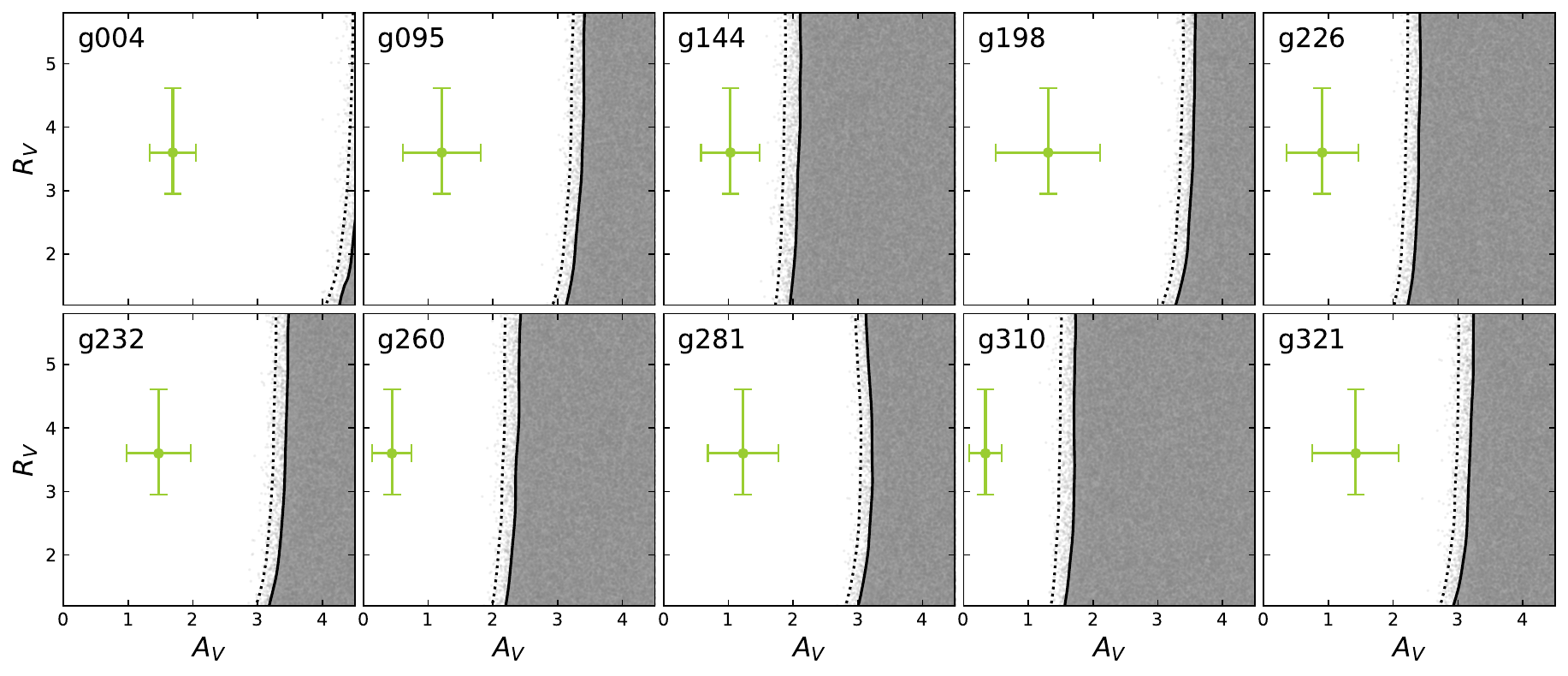}
        \caption{Our samples and the contour of $A_V$ and $R_V$ values that makes each galaxy clip the limiting magnitude.}
        \label{fig:limiting_magnitudes}
    \end{figure*}

    The results of this test are shown in Figure~\ref{fig:limiting_magnitudes}. We measure the density of ``unobservable'' ($A_V$, $R_V$) samples and define $50\%$ density (unobservable half of the time under bootstrapping) as the threshold to mark the ``unobservable'' region. We draw the $1\sigma$ region (possibly unobservable) at the $16\%$ density. Our results show that all of our galaxies are $\gtrsim2\sigma$ away from interfering with the grayed region regardless of the $R_V$ value. From this test, we conclude that it is unlikely that our measurement of $R_V$ is biased by the sampling of background galaxies.
    
    \begin{figure}[h]
        \centering
        \includegraphics[width=\linewidth]{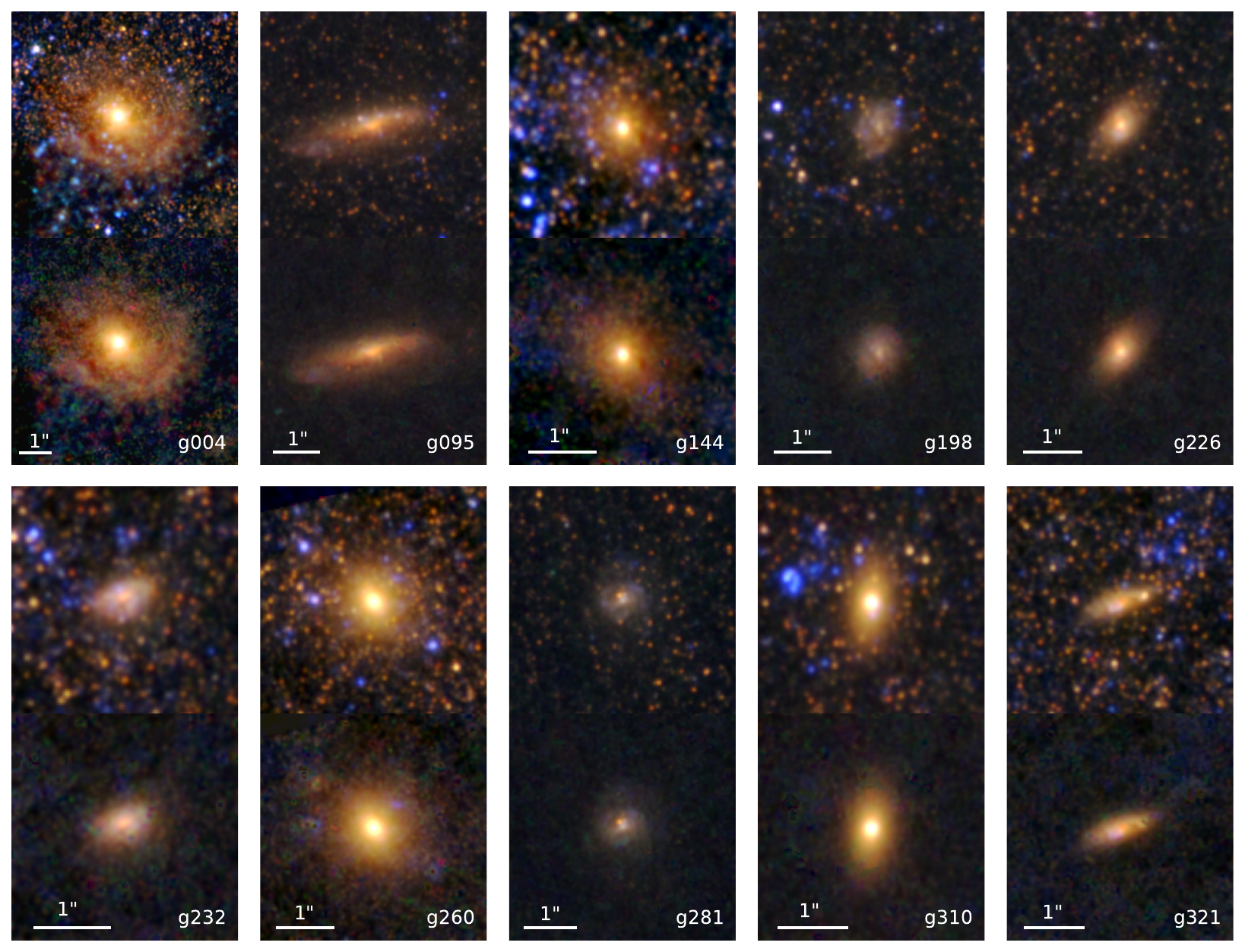}
        \caption{False-color images of our target background galaxies. The scaling is fixed across all galaxies for each filter to provide a visual comparison for each galaxy's color and brightness. The top panel of each galaxy shows an unaltered image, and the bottom panel shows the \sphot-processed images at the same flux scale.}
        \label{fig:RGB_cutouts}
    \end{figure}
    
    \begin{figure}[h!]
        \centering
        \includegraphics[width=\linewidth]{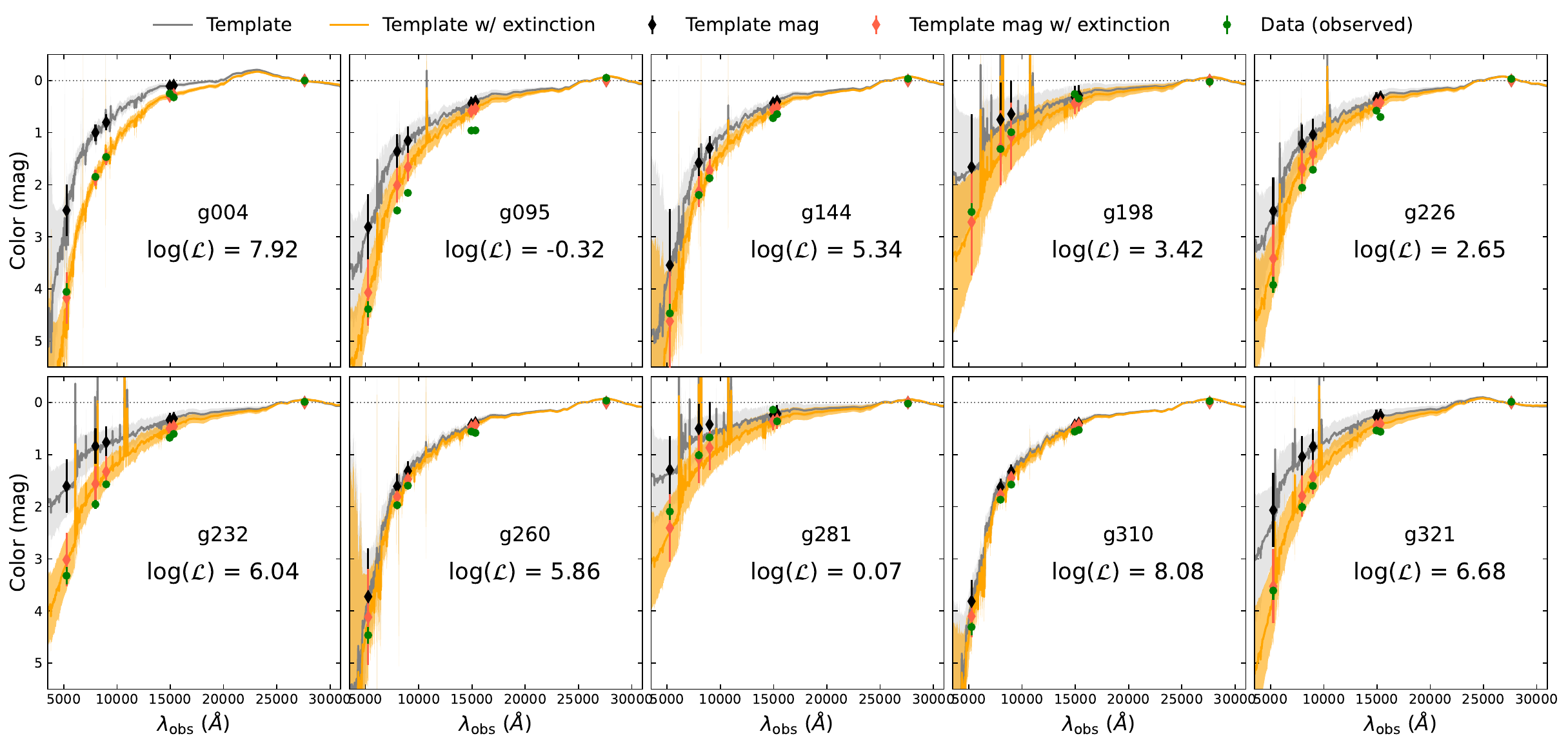}
        \caption{Individual SED of our background galaxies and estimated intrinsic SED. Gray curves indicate the combined template and its uncertainty, with black dots representing the bandpass-filtered synthetic photometry. Green dots represent the measured SED. The combined templates with the best-fit extinction applied are shown as the orange curves.}
        \label{fig:individual_sed}
    \end{figure}
    
    \begin{figure}[h!]
        \centering
        \includegraphics[width=\linewidth]{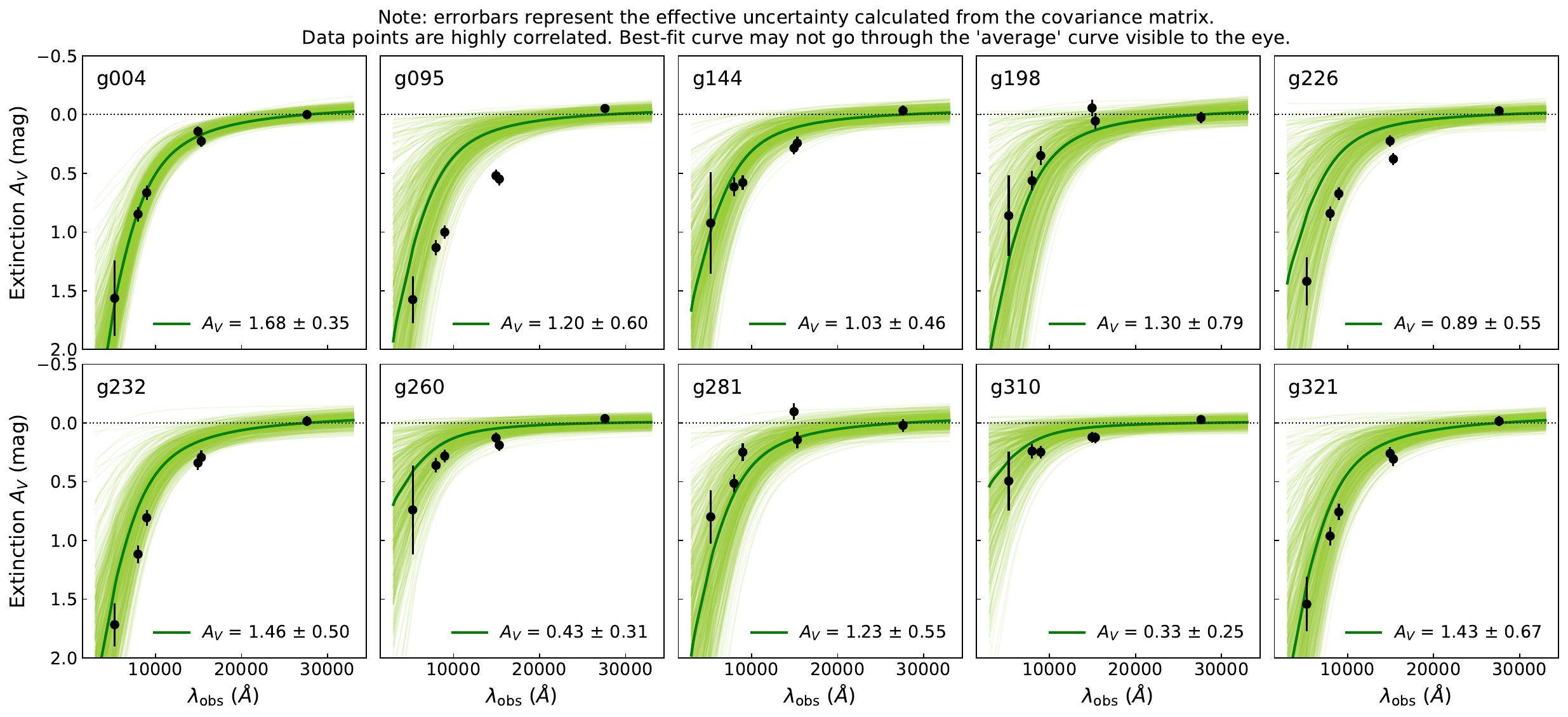}
        \caption{Individual extinction measured and the best-fit extinction curves. Black dots represent the difference between the estimated intrinsic color and the observed color at each band, and the green curve represents the best-fit extinction curve, as well as individual samples from MCMC. Note that all of these data are fit \textit{simultaneously} and there is a significant covariance between photometric data points: the ``best fit'' for each galaxy may not go through the center of all data points.}
        \label{fig:individual_ext_curves}
    \end{figure}

    \begin{figure}[h!]
        \centering
        \includegraphics[width=\linewidth]{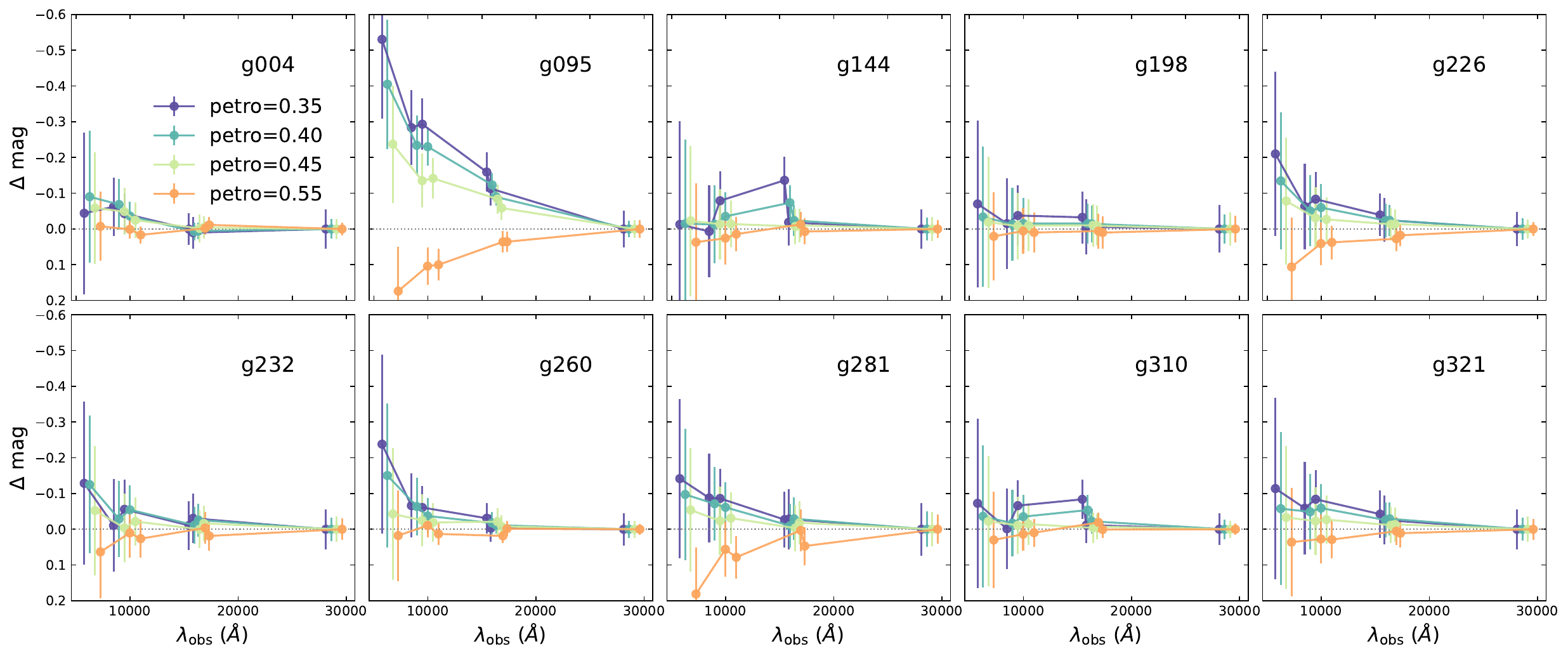}
        \caption{The comparison of measured SEDs with respect to our baseline photometry, measured at different aperture sizes. Larger Petrosian values correspond to the smaller aperture sizes, and the small change in the observed SED, due to the color gradient of galaxies, is visible. All but two galaxies have such trends well within the uncertainty size. The effect of aperture size is discussed in Sec.~\ref{sec:discussion}.}
        \label{fig:aperture_effect}
    \end{figure}

\bibliography{main}{}
\bibliographystyle{aasjournal}

\end{document}